\documentclass[aps,prl,groupedaddress]{revtex4}
\usepackage[latin1]{inputenc}
\usepackage{amsmath}
\usepackage{verbatim}
\usepackage{amssymb}
\usepackage{bbm}
\usepackage{txfonts}
\usepackage{graphicx}
\usepackage{amsmath, amssymb,mathtools}

\begin{document}
\title[Robust transformations of firing patterns for neural networks]{Robust transformations of firing patterns for neural networks} 
\author{Karlis Kanders$^{1}$, Tom Lorimer$^{1}$, Yoko Uwate$^{2}$, Willi-Hans Steeb$^{3}$, 
Ruedi Stoop$^{1,3,4,*}$}
\address{Institute of Neuroinformatics and Institute for Computational Science, University and ETH Z{\"u}rich\\ Irchel Campus, Winterthurerstr. 190, 8057 Z{\"u}rich, Switzerland\\
Email: ruedi@ini.phys.ethz.ch\\
$^{1}$ Institute of Neuroinformatics, University of Zurich and ETH Zurich, Zurich, Switzerland \\
$^{2}$ Electrical and Electronic Engineering, Tokushima University, Japan\\
$^{3}$ School of Computation, University of Johannesburg, Republic of South Africa\\
$^{4}$ Institute for Computational Science, University of Zurich, Zurich, Switzerland }
\def\corrAuthor{Ruedi Stoop}

\def\corrEmail{ruedi@ini.phys.ethz.ch}

\begin{abstract}
As a promising computational paradigm, occurrence of critical states in artificial and biological neural networks has attracted wide-spread attention. An often-made explicit or implicit assumption is that one single critical state is responsible for two separate notions of criticality (avalanche criticality and dynamical edge of chaos criticality). Previously, we provided an isolated counter-example for co-occurrence. Here, we reveal a persistent paradigm of structural transitions that such networks undergo, as the overall connectivity strength is varied over its biologically meaningful range. Among these transitions, only one avalanche critical point emerges, with edge of chaos failing to co-occur. Our observations are based on ensembles of networks obtained from variations of network configuration and their neurons. This suggests that not only non-coincidence of criticality, but also the persistent paradigm of network structural changes in function of the overall connectivity strength, could be generic features of a large class of biological neural networks.  

\end{abstract}
\maketitle

\section{Introduction}

Ensembles of interacting particles (such as spin systems or neural networks) are generally described by means of a thermodynamic function. At points of nonanalytic behavior of the thermodynamic function, the function and the system are said to be at a phase transition. More precisely, using the terminology of Ehrenfest, we are at a  $k^{th}$ order phase transition if the $k^{th}$ derivative of the Gibbs potential fails to exist. A newer terminology only distinguishes first and second order transitions, where the latter class encompasses all higher $k>2$ order transitions. At a `critical point' of the thermodynamic function, we are at a second order phase transition.

The melting of ice or the boiling of water, are the showcase physics examples of a first order phase transition. Here, the temperature of the system stays constant as heat is added: the system is in a "mixed-phase regime", in which some parts of the system have completed the transition while others have not. Second-order phase transitions are more subtle. In the neighborhood of the transition, they exhibit  a diverging susceptibility and a power-law divergence of correlations; the well-known physical example is the ferromagnetic transition. Phase transitions often involve breaking of symmetry: The cooling of a fluid into a crystalline solid breaks continuous translation symmetry, illustrating that the high-temperature phase  typically contains more symmetries than the low-temperature phase (infinite-order phase transitions break no symmetries, as exemplified by the two-dimensional XY model).
Thus, the precise nature of phase-transitions strongly depends on model particularities; the $q$-state Potts model \cite{Potts}, e.g., has for all $q \geq 3$ in 3D a first-order transition, but in 2D a second-order transition for $q \leq 4$ and a first-order transition for all $q \geq 5$ \cite{Wu1}.
Taking magnetization $m$ as the order parameter, because of the two phases present, first order transitions are characterized by a finite jump at the transition point (from a positive value to zero), of the specific heat (or the susceptibility). In second order transitions, the magnetization goes to zero in a power-law manner, with the specific heat or the susceptibility developing power-law poles. Both power laws are associated with (generally distinct) critical exponents. At a second order transition, also the correlation length $\xi$ diverges (with generally yet another power-law exponent). Altogether, denoting by $T_c$ the critical temperature, we have
 \begin{eqnarray}
 \xi=\xi_{o\pm}+ \mid 1-T/T_c\mid ^{-\nu}+... , \\
C=C_{reg} +C_o \mid 1-T/T_c\mid ^{-\alpha}+... , \\
m = m_o \,(1- T/T_c)^{\,\,\,\,\beta}+... , \\
\chi=\chi_o\, \mid 1-T/T_c\mid ^{-\gamma}+...  \,,
\end{eqnarray}
where $\xi$ is the correlation length. Generally, the approaches to the critical point from both sides differ, $\xi_{o+}\neq \xi_{o-}$); $C$ is the specific heat, $m$ the `magnetization', and $\chi$ the susceptibility of the system. These relations define different `critical exponents' $\nu, \alpha, \beta, \gamma$.  

In {\em finite size} systems, such as we shall deal with in our paper (as in any numerical simulation or  experimental contexts), the correlation length cannot diverge, and the divergences are shifted and mollified. In the vicinity of the critical temperature $T_c$, the linear system size $L$ takes the role of $\xi$, so that the scaling formulas 
\begin{equation}
C= C_{reg} +\alpha L^{\alpha/\nu} +..., \,\,\,\,\,  \\
 m \propto L^{-\beta/\nu}+... , \,\,\,\,\, \\
\chi\propto L^{\gamma/\nu} +.... 
\end{equation}
emerge. If the scaling variable $x = (1 - T /T_c) L^{1/\nu}$ is kept fixed, this reveals a deeper insight: The susceptibility of the system, e.g., scales as $\chi(T, L) = L^{\gamma/\nu} f(x)$, so that if $\chi(T, L)/ L^{\gamma/\nu}$ is plotted vs. scaling variable $x$, the data for different $T$ and $L$ are predicted to fall onto the scaling function $f(x)$. The corresponding observation for first-order phase transitions is that the original delta-like singularities from phase coexistence, are smeared \cite{f1,f2,f3}. 
They are replaced by narrow peaks, the height and width of which grow proportional to system volume and volume$^{-1}$, respectively \cite{f4,f5,f6}.
We finally note that, upon varying external parameters (like the magnetic field, node or interaction composition), phase transition points can generalize to transition lines, including multicritical points
 \cite{stoopdiffusion1,stoopdiffusion2,stoopdiffusion3}.

In the identification of real-world realization of criticality in biology, one deals with a number of problems, each of which can be expected to prohibit proper critical transitions. In addition to finite size (that biology shares with all other real-world manifestations of this concept), we generally deal with open systems (they react upon external input) with complicated node dynamics. Moreover, biological systems are often characterized by several levels of hierarchies, which puts them even farther away from anything like the thermodynamic limit. Generally, at each level of the hierarchy, they use `whatever architecture works' (a well-known example is the co-existence of chemical and of electrical synapses in the brain). Only rarely, the implementation of one well-defined fundamental solution principle can be shown to persist in an exclusive manner (such as in the mammalian hearing organ, the cochlea \cite{Lorimerscirep14}, but even in this case biology can be seen to exploit local, non-scaling, optimization).

\section{Criticality in cortical network models}

Spins can be seen as the simplest model of neuronal firing, where `spin up' codes, e.g., for an excited (firing) neuron. From this, a natural suggestion emerges of what kind of transitions to expect in neural networks. The brain, that has been a focus of recent criticality approaches (for a literature overview see \cite{Gross14}), exhibits, however, a multitude of distinguished architectural levels, which is not foreseen in the models from physics. We will focus on a neural network type motivated by the brain's cortical column \cite{stoopprl2013}. This physiological structure is widely conceived as a fundamental architectural entity above the single neuron level (occasionally, this view is confronted with scepticism \cite{Rakic}, as striking divergences between architectural and dynamical characterizations can be exhibited \cite{Jordi2013}). Beyond the columnar size, additional architectural characteristics become relevant \cite{stoopprl2013} and close-to biology columnar modeling will also contain properties reflecting the connectivity to higher hierarchy levels. To arrive at the thermodynamic limit would therefore require to massively remove essential biology from a detailed model of the cortical column.
The {\it approximative scaling approach} followed here (e.g. \cite{Gross14}) solves this problem by taking into consideration scaling only across the range directly related to the scale of the model (here: the column size \cite{stoopprl2013}). The cortical hierarchy can display properties ascribed to critical systems nonetheless, if each hierarchy level optimizes its behavior at its own level. This requires using the principles of thermodynamics in an slightly more generalized manner (another example of such an extension is thermodynamic formalism of dynamical systems \cite{stoopprl2016}), but may provide a consistent framework for revealing how composite complex entities optimize their information processing. In this generalized sense, we propose for the following to call states `critical', if they sufficiently share (topologically or dynamically) the statistical properties of finite size states of systems close to a (provable) critical point. 

Particular importance has been attributed in this context to critical points, because these points could be associated with enhanced computational power. This idea has been exploited into different directions. Enhanced computational power is often seen as a kind of `increased flexibility of response in dependence of stimulation', combined with the `ability to best accommodate temporal patterns' ({\em `edge of chaos criticality'}, e.g. Ref. \cite{Langton1990}). Focusing on the ability of how to best accommodate complex input within the network, the view has been that at criticality, computation would be highest. 
Similarly, enhanced computational power has been attributed to situations where avalanches of neuronal firing lack a characteristic size (number of firing events triggered) or duration (time across which such avalanches extend), i.e., if the corresponding distributions have a power-law form \cite{Beggs2003}. We will refer to this situation as  {\em`avalanche criticality'}. Power law distributions have been found in a vast variety of experimental settings using electrophysiological recordings in vitro \cite{Beggs2003,Mazzoni2007, Pasquale2008} and in vivo \cite{Petermann2009,Hahn2010}, as well as EEG recordings \cite{Allegrini2010, Palva2013} and fMRI \cite{Fraiman2009,Tagliazucchi2012}.

For the validation of avalanche criticality in biological neural networks, a number of features have emerged as instrumental: (1) a close relationship between the average avalanche size and lifetime power laws \cite{Friedman2012, Shew2015}, (2) resilience of the power law behavior against temporal binning with different bin widths \cite{Beggs2003, Priesemann2014}, (3) fulfillment of a mathematical relationship between the critical exponents, (4) self-similarity of avalanches indicated by a collapse onto a single shape after proper rescaling \cite{Friedman2012,Sethna2001}. 
The ubiquity of the observations of avalanche criticality have encouraged a hypothesis that biological neural networks self-organize towards critical states: Avalanche criticality has indeed been reported as the final state in the development of cultured neurons in-vitro \cite{Tetzlaff2010}, as well as the result of sensory adaptation in living brains \cite{Shew2015}. As the mechanism for self-organization towards the critical state, activity-dependent synaptic plasticity has been proposed \cite{Levina2007,Meisel2009}.
However, as yet it has not been possible to link avalanche criticality to a particular statistical mechanics universality class \cite{Priesemann2014}. Whereas in the first experimental studies of avalanche size distributions a power-law exponent 1.5 was observed  \cite{Beggs2003}, indicating the possibility of the underlying mechanism to be the critical branching process, later studies have suggested a rather broad range of exponents \cite{Shew2015,Tetzlaff2010,Levina2017}, sizing up even to 2.6.\\

The two notions of criticality have been linked in terms of the so-called `criticality hypothesis', stating that brains operate at criticality to `optimize computation' \cite{Beggs2008, Gross14}, implying that experiments showing evidence of avalanche criticality, in some way support the hypothesis that living systems preferably operate at edge-of-chaos \cite{Beggs2008}. Motivated by this, avalanche criticality has been described as something akin to a state between `total randomness and boring order' \cite{Beggs2008}. Few investigations have subsequently indeed found co-occurrence of dynamical criticality and emergence of fingerprints of avalanche criticality in neural networks \cite{Haldeman2005, Magnasco2009}. These models, however, were built upon abstract neuron models that do not encompass the lower-level details of neuronal dynamics, which could have a profound influence on the overall network behavior. 
More recent approaches addressing this topic on different scales of the cortical hierarchy are, at the macroscopic level, self-organized bistability-induced power laws \cite{Munoz}, at a more detailed level, spin systems on the Human brain connectome \cite{Marinazzo}, and activity-induced power-laws on neuronal networks of simplified neuron models \cite{Lucilla}.

Edge of chaos based on more realistic spiking neuron models has been investigated in the context of computation in recurrent networks \cite{Legenstein2007}. In this case, computational benefits have been claimed by providing a whole spectrum of states representing integrated inputs \cite{Legenstein2007, Bertschinger2004}. In these works, the presence or absence of avalanche criticality has, however, not been addressed. From a similar perspective, also avalanche criticality has been attributed to maximize information transmission and capacity \cite{Shew2011}, where the focus was on the high number of metastable states \cite{Haldeman2005} that might serve a substrate for information storage. 

While our own interpretation of computation deviates somewhat from these views (we feel that the fundamental principle that must be attached to a notion of computation is the process of information destruction \cite{Stoop2004, stoopprl2016}, which is missing in these frameworks), we fully concur with the importance of understanding the basis, occurrence and function of critical states.

\section{A close-to-biology neural network framework}

Our example \cite{Kanderschaos} contradicting necessary coincidence of avalanche and edge-of-chaos criticality, was based on a meso-scale model of a cortical column. Emphasis was put onto conserving as much as possible of the example's network structure and firing patterns, whereas more microscopical aspects, such as the modeling of internal neuronal currents, were abandoned. This has as a consequence this modeling is confined to columnar size; larger scales are not within the model, in particular, since the whole cortex is highly hierarchic, with largely non-selfsimilar levels. For this reason, the thermodynamic limit cannot be expected to be attainable by simply scaling up the columnar model.
Our present investigations can be seen as a continuation of the former work; the model description therefore closely follows Ref. \cite{Kanderschaos}. The network is built on a network graph, with neurons of the Rulkov model (see below) as the nodes. These neurons are set below firing threshold \cite{Kanderschaos}; most of the time most biological neurons are below firing threshold. Nodes are brought to activity - additively or alternatively - by means of external input, or by naturally persistently firing neurons, so-called `leader neurons'  \cite{Eckmann2008}  (also called  `nucleation sites' \cite{Jordi2013}). Our use of the term is, however, more general by taking it out of a bursting context; bursting may, but does not need to happen in our setting. 
The external noisy input we chose as nonspecific as possible, and to have a much inferior effect compared to that by the leader neuron(s) (see below). We note, however, that for a large range of combinations of the activities of the two sources, networks exhibit similar behavior. 
External noisy input accounts for `dynamical randomness', a feature of realistic neuronal firing (e.g., synaptic noise from spontaneous vesicle release). To model this, we implemented for each neuron an individual excitatory Poisson spike train input. This input was represented by a spike variable $\xi^{ext(i)}_n$ having value 1 when the $i$-th neuron receives an external spike and 0 otherwise
\begin{equation}
\xi^{ext(i)}_{n+1} =
    \begin{dcases}
  1 & p < p^{ext}, \\
  0 & p \geq p^{ext},
    \end{dcases}
\end{equation}
where $p$ is a random number drawn at each time step from a uniform distribution in the open interval (0,1). Choosing $p^{ext} = 6 \cdot 10^{-4}$ renders the external input temporally sparse.  

Leader neurons represent entry points of coherent external (e.g., sensory) input. Such input can best be implemented by putting excitatory neurons above the firing threshold (by a suitable increase of the neuron's excitability parameter $\sigma$, see the Rulkov model below). In this case, this neuron fires in a subtly chaotic manner. This firing behavior is then modified further by means of recurrent connections with the network (Figs. 1a), 2b), c)). For the case reported here, a single intrinsically firing neuron was implemented (as shown in Fig. 1a)), but, if desired, several of such elements can be used.

Apart from this, the network is built on the well-known 1: 4 part composition of inhibitory vs. excitatory neurons (cf. Fig. 1a)), where inhibitory synapses are generally much stronger than excitatory synapses (by a factor of around 3 in our model). Our coarse-grained connectivity graph was built upon a connectivity probability of $p_c=0.04$; the edges between network nodes are directed. Each neuron $i$ in the network has a number $k$ of `presynaptic', i.e. impinging, neurons; the full relationship is defined by the network's weight matrix $w_{ij}$, where $i,j=1,\ldots, N$, where $N$ is the number of neurons in the network (Fig. 1b)).
Specifically, the number of presynaptic excitatory and inhibitory nodes was set to $N^{pre}_{ex} =4 \simeq N_{ex} \,p_c $ and $N^{pre}_{inh} =1\simeq  N_{inh}\, p_c $ (rounded to the nearest integer). For every neuron, $N^{pre}_{ex}$  and $N^{pre}_{inh}$ nodes were randomly selected as presynaptic neighbors, from the pool of excitatory and inhibitory neurons, respectively; self-connections were eliminated. By construction, network nodes thus had an in-degree $k_{in}$ of 5 or 4 (the latter case due to the elimination of self-connections). In this simple and controllable way, heterogeneity is introduced in the network, where  out-degrees vary more strongly than in-degrees and where neurons with high out-degree dominate the network activity. Connections between neurons modeled in this way represent a mesoscale implementation of the synaptic coupling between biological neurons. The number of neurons in this network was usually $N = 128$, with $N_{ex}=102 \simeq  0.8 N$ excitatory neurons and $N_{inh} = 26 \simeq 0.2 N$ inhibitory neurons (rounding to the nearest integer).
Network size, topology and input structure approximate a cortical column \cite{stoopprl2013}; additionally, this offers a good trade off between obtaining enough statistics for the avalanche size distributions, at reasonable computational expenses for extracting in parallel other important network characteristics, such as network Lyapunov exponents. 

\begin{figure}[]
\includegraphics[width=0.85\linewidth]{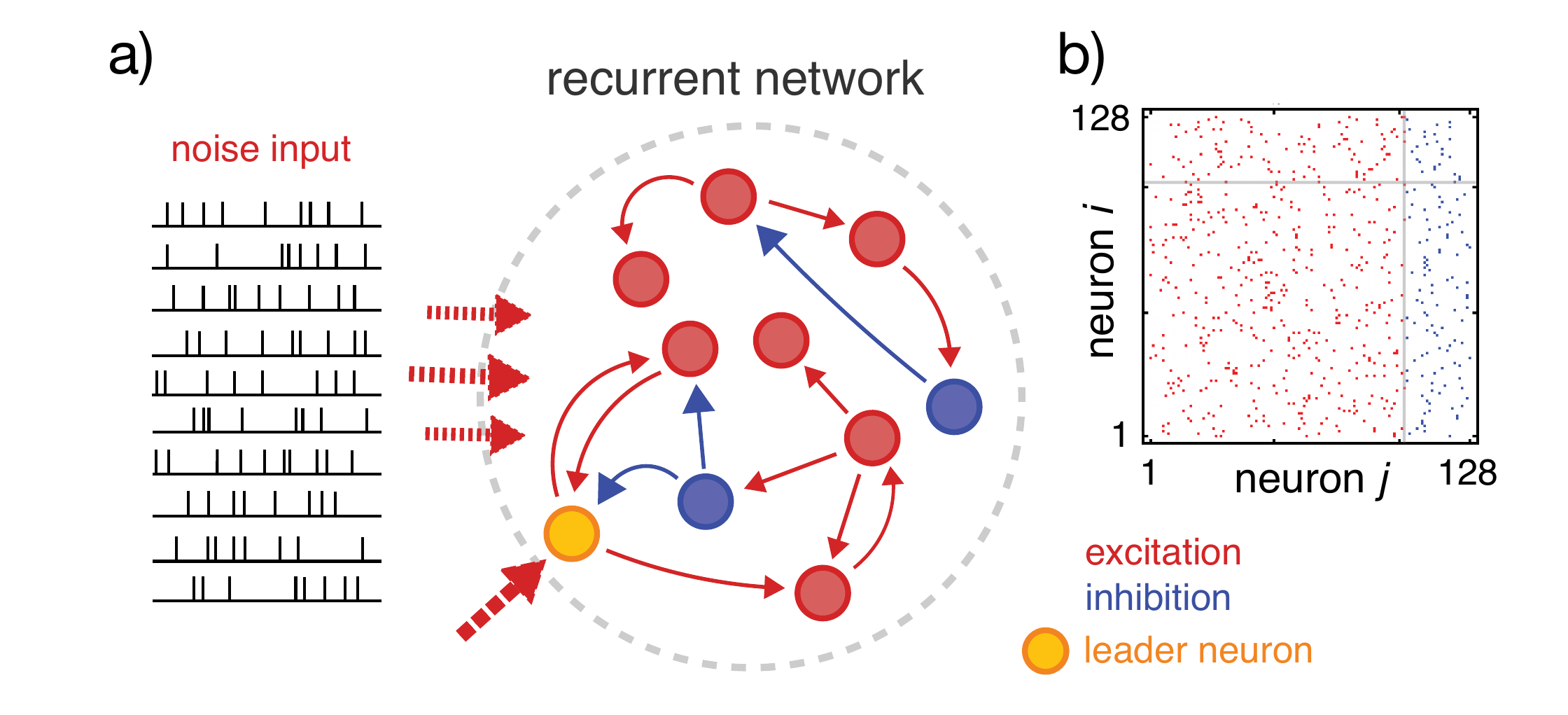}
\caption{Network architecture \cite{Kanderschaos}: a) Recurrent network with excitatory neurons (red circles), inhibitory neurons (blue circles) and one intrinsically firing excitatory neuron (yellow circle) representing entrant sensory stimulation. Each neuron also receives noisy excitatory input, from a Poisson spike train. b) Weight matrix $w$ of the network, with excitatory connections (red) and inhibitory connections (blue). Grey lines are for the last excitatory neuron (index $j=102$). }
\label{modeltopology}
\end{figure}

\begin{figure}[]
\includegraphics[width=0.75\linewidth]{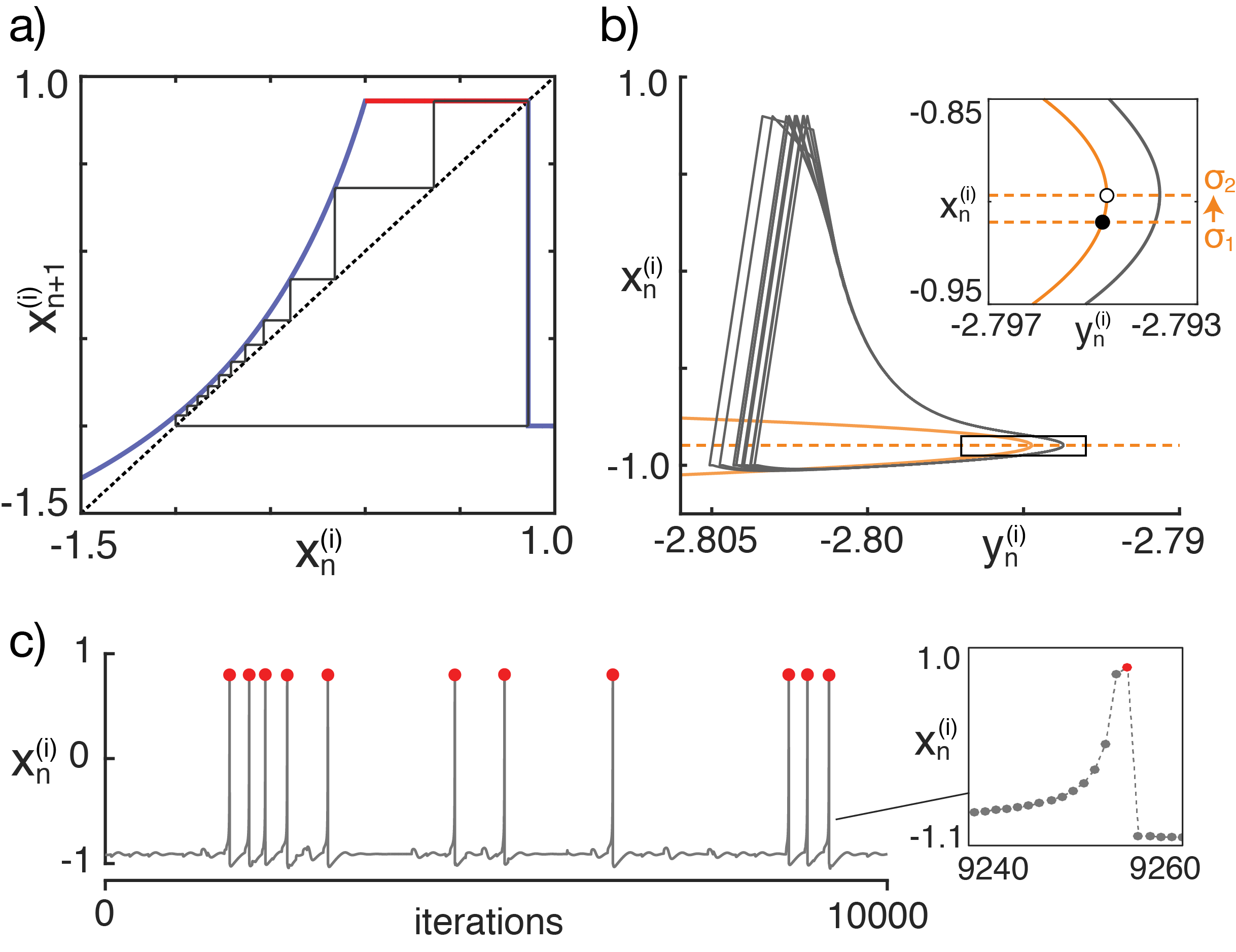}
\caption{Network node dynamics \cite{Kanderschaos}:  a) Iterates of the Rulkov map for fixed $y^{(i)}=-2.74$ (particular choice for display reasons). b) Full trajectory for the intrinsically firing Rulkov neuron (solid gray line) in the absence of external input. Orange line: $x^{(i)}_n$ nullcline $y^{(i)}_n = x^{(i)}_n - \psi / (1-x^{(i)}_n)$; dashed orange line :  $y^{(i)}_n$ nullcline $x_n^{(i)} = -1 + \sigma$. Increase of $\sigma$ from $\sigma_1=0.09$ to $\sigma_2=0.103$ shifts the $y^{(i)}_n$ nullcline vertically (arrow) and changes the map's central fixed point from stable (filled circle) to unstable (empty circle). c) Firing pattern of a typical network-embedded neuron ($\sigma = \sigma_1$). Red circles denote spike events (i.e., $\xi^{(i)}_n = 1$). Right hand side box: Magnification of the spike-generating discrete dynamics.}
\label{modeldynamics}
\end{figure}

As was mentioned above, to model the dynamics of a network neuron labeled with index $i=1,\ldots,128$,
we use Rulkov's two-dimensional iterative map \cite{Rulkov2002, Rulkov2004}
\begin{subequations}
\label{Rulkov}
\begin{align}
    x^{(i)}_{n+1} &= 
    \begin{dcases}
    \frac{\psi}{1-x^{(i)}_{n}} + u_n^{(i)} & x^{(i)}_n\leq 0,\\
    \psi + u_n^{(i)} & 0 < x^{(i)}_n < \psi + u_n^{(i)} \wedge x^{(i)}_{n-1}\leq 0,\\
    -1 & x^{(i)}_n \geq \psi + u_n^{(i)} \lor x^{(i)}_{n-1} > 0,
    \end{dcases}\\ \nonumber
y^{(i)}_{n+1} &= y^{(i)}_{n} - \mu(1+x^{(i)}_{n}) + \mu\sigma + \mu I^{(i)}_n,\\
\end{align}
\end{subequations}
where the iteration step is denoted by index $n$ and where $x^{(i)}_n$ models the neuron's membrane potential. $y^{(i)}_n$ describes a regulatory subsystem able to turn the firing on and off (Figs. \ref{modeldynamics}a), b)). $u_n^{(i)} =  y^{(i)}_{n} + \beta I^{(i)}_n$; $ I^{(i)}_n$ describes the postsynaptic input to neuron $i$ (see below). Parameter $\sigma$ controls the state of the map (Fig. \ref{modeldynamics}b)), where $\sigma = 2 - \sqrt{\psi/(1-\mu)}$ at the bifurcation point. The parameter values for excitatory and inhibitory neurons are randomly chosen from Gaussians around the values $\psi = 3.6$, $\mu = 0.001$, $\sigma=0.09$, $\beta = 0.133$ (see Appendix for details). For these values, around $\sigma=0.101684$, Rulkov's map undergoes a subcritical Neimark-Sacker bifurcation from silent to firing behavior, corresponding to a Class II neuron behavior \cite{Prescott2008}. 
Rulkov neurons can reproduce essentially all experimentally observed firing patterns \cite{Rulkov2004}, and even finer neurobiological details \cite{Kanders2015}. By this feature, our network model distinguishes substantially from the previous efforts of linking avalanche and edge of chaos criticality based on probabilistic binary units \cite{Haldeman2005} or abstract `rate' neurons \cite{Magnasco2009}.
These characteristics of the model are collected in Figs. \ref{modeltopology}, \ref{modeldynamics}.

Rulkov neurons receive spike input from other neurons via synapses (attached to the message-receiving neuron) as follows. A firing variable $\xi^{(i)}_n$ carries the value 1 if at iteration $n$, neuron $i$ generated a firing event (spike), and a value 0 otherwise:
\begin{equation}
\xi^{(i)}_{n+1} =
    \begin{dcases}
  1 & 0 < x^{(i)}_n < \psi + u_n^{(i)} \wedge x^{(i)}_{n-1}\leq 0. \\ 
  0 & \text{otherwise}.
    \end{dcases}
\end{equation}
I.e., neuron $i$ is firing at iteration $n$,  if $x^{(i)}_n$ attains the maximum value of the map (red horizontal line in Fig. \ref{modeldynamics}a)).
Synapses have their own dynamics that can be modeled by an exponential decay and step-like increase upon presynaptic spike events as
\begin{equation}
    I^{(i)}_{n+1} = \, \,\eta I^{(i)}_{n} + W \bigg( \sum^{N_{ex}}_{j=1} w_{ij}(x^{ex}_{rp} - x^{(i)}_n) \xi^{(j)}_n 
     + \sum^{N}_{j=N_{ex}+1} w_{ij}(x^{inh}_{rp} - x^{(i)}_n) \xi^{(j)}_n 
     + w_{ext} (x^{ex}_{rp} - x^{(i)}_n) \xi^{ext(i)}_n \bigg).
\label{eq_synapse}
\end{equation}
$\eta$ controls the decay rate of the synaptic current and $x^{ex}_{rp}$ and $x^{inh}_{rp}$ are the reversal potentials of excitatory and inhibitory synapses, respectively. In Eq. \ref{eq_synapse}, the corresponding contribution vanishes, if there is no connection between the neurons $i$ and $j$ ($ w_{ij} = 0$, where $i$ is the index of the postsynaptic neuron and $j$ is the index of the presynaptic neuron (Fig. \ref{modeltopology}b)), or if there was no presynaptic spike event ($\xi^{(j)}_n = 0$). The decay parameter is chosen as $\eta = 0.75$, the reversal potentials as $x_{rp}^{ex} = 0$ and $x_{rp}^{inh} = -1.1$, respectively, and the external input weight as $w_{ext} = 0.6$. Excitatory connections have a weight of $w_{ij} = 0.6$, inhibitory connections have the tripled weight ($w_{ij} = 1.8$). By joining the effect by  $\sigma$, $I_n^{(i)}$ can push intrinsically silent neurons into firing (Fig. \ref{modeldynamics}). 
In Eq. \ref{eq_synapse}, $W$ is a connectivity-scaling factor. Increasing $W$ enhances the coupling among the neurons without changing architecture otherwise. 

\section{Relevance of network model and results }

In our previous work \cite{Kanderschaos}, we exhibited an example contradicting the assumption of the necessary coincidence of the two notions of criticality.  For the counter example, three parameter values from the close vicinity of the avalanche transition point were investigated, from the avalanche criticality and Lyapunov exponent point of view. Outside this neighborhood, lots of structural and dynamical changes could, however, occur. Multiple (in fact: a countable infinity of) coincidences of criticality could emerge, which would qualify our earlier findings. To eliminate this terra incognita, we performed extensive computational explorations, until a detailed reliable overview of the behavior of the system was obtained. The insight conveyed now, is the extension from a counter example (a zero measure statement) to showing that such counter-examples will be the generic case (i.e., non-coincidence has a non-zero measure in the space of systems). Generic refers here to our model class that we consider to be both sufficiently detailed  and yet sufficiently general, regarding the properties of the biological example (see below). To work out the generic system behavior, we characterized the network state by avalanche size- and life-time distributions, largest Lyapunov exponent, Kolmogorov-Sinai entropy, firing rate, and synchronization measure.

On top of the topological structure given, the network's metric activity is also determined by the strength of the individual connections and the more dynamical aspects of firing. We therefore assess network firing behavior based on ensembles of network realizations that we individually construct on distributions of biologically plausible parameter regions. Neurons providing the node dynamics were similarly constructed (see Appendix for parameter distributions). 
Rulkov neurons are an extremely well-tested \cite{Kanders2015}, mesoscale, phenomenological model of realistic neuronal behavior. In addition to reproducing almost any temporal firing pattern, Rulkov neurons have been shown to reliably exhibit very specific features of realistic neuronal firing, such as the phase response curves of physiologically measured biological neurons of different kinds \cite{Kanders2015}. By retaining in this sense a large portion of realistic neuronal dynamics, they are in optimal harmony with a tractable mesoscale neuronal network description. Our approach provides, moreover and more generally, valuable guidelines for developing reliable models of neural network dynamics that can be used for statistical approaches.

On this description level, we will scrutinize what kinds of transitions between phases of neuronal firing behavior should be expected for such systems. In this framework, neurons that are generally below the firing threshold are occasionally activated by noise, by the activity provided by leader neurons, or by generated network activity. Leader neurons were constructed using the same model, but endowed with an augmented excitability parameter $\sigma$ (see Appendix), which leads them to spike intrinsically in a chaotic manner. This property was desired as the role of leader neurons was to represent structured, but temporally changing input from outside (e.g., sensory input into the network). Preferably, such input should neither be periodic, nor sporadic, nor pure noise, but should have chaotic dynamics. 
When inserted into a network, a chaotic leader neuron may, however, become regularized or silenced, by the interaction with other neurons. Already the coupling of non-identical simple systems can exhibit behavior virtually unpredictable from the properties of isolated contributing systems (this has been a major  focus of Nonlinear Dynamics over decades). A chain of arguments like `chaotic leader neuron $->$ chaotic subsystem $->$ no chance of coincidence of transitions' would therefore be premature; the outcome of such a coupling depends in a complicated manner on the particularities of the contributing systems and their mutual relationship. After sampling the network and the neuronal properties, the overall coupling strength parameter $W$ is the only free parameter left to scan the firing behaviors produced by such networks. Increasing $W$ increases the coupling among the neurons, without changing architecture otherwise. By changing $W$, diverse activity states can be accessed, based on an ensemble of network topologies that provide a statistical basis. 

The extrapolation from the model to the behavior of biological structures such as cortical columns, requests a model that is extremely close to biological reality. To support this, we took great care to implement as much as was possible biological knowledge, rather than looking at simpler models that would be easier to treat. Our explorations will show that at moderate excitations, edge-of-chaos is excluded and does therefore not concur with avalanche criticality.

\section{Network simulations}

Despite their random parameter configurations underlying the individual systems, our simulations yielded very coherent results, exhibited by a very small variability of the characterizing measures around the behavior exhibited by characteristic networks (i.e., networks at central parameter values). A single simulation covered $5 \cdot 10^5$ time steps, of which the first 5000 steps were discarded. For our ensemble, at each value of $W$, we chose the first 25-50 simulations that exhibited a requested level of activity (at the critical point we occasionally increased this number to enable more precise statistical characterizations). The precise requirement was that the average inter-event interval $\langle IEI \rangle$ (i.e., the average time between two subsequent spikes in the network) of a simulation run should fall into the interval $\mu_{\langle IEI \rangle} \pm \frac{\sigma_{\langle IEI \rangle}}{1.5}$, where $\mu_{\langle IEI \rangle}$ and $\sigma_{\langle IEI \rangle}$ are the mean and standard deviation of the $\langle IEI \rangle$ distribution across all network simulations at a particular $W$. Such sampling ensures that results from typical network realizations are looked at, and it permits the pooling of the results from different simulations. Except for the in-degree that, however, was checked to have no significant influence in what follows, 
all network and neuronal parameters were chosen randomly from biologically plausible intervals.
The overall coupling strength $W\in[0,1]$ provides therefore the relevant handle to explore the network properties. 

\begin{figure}[h]
\includegraphics[width=0.75\linewidth]{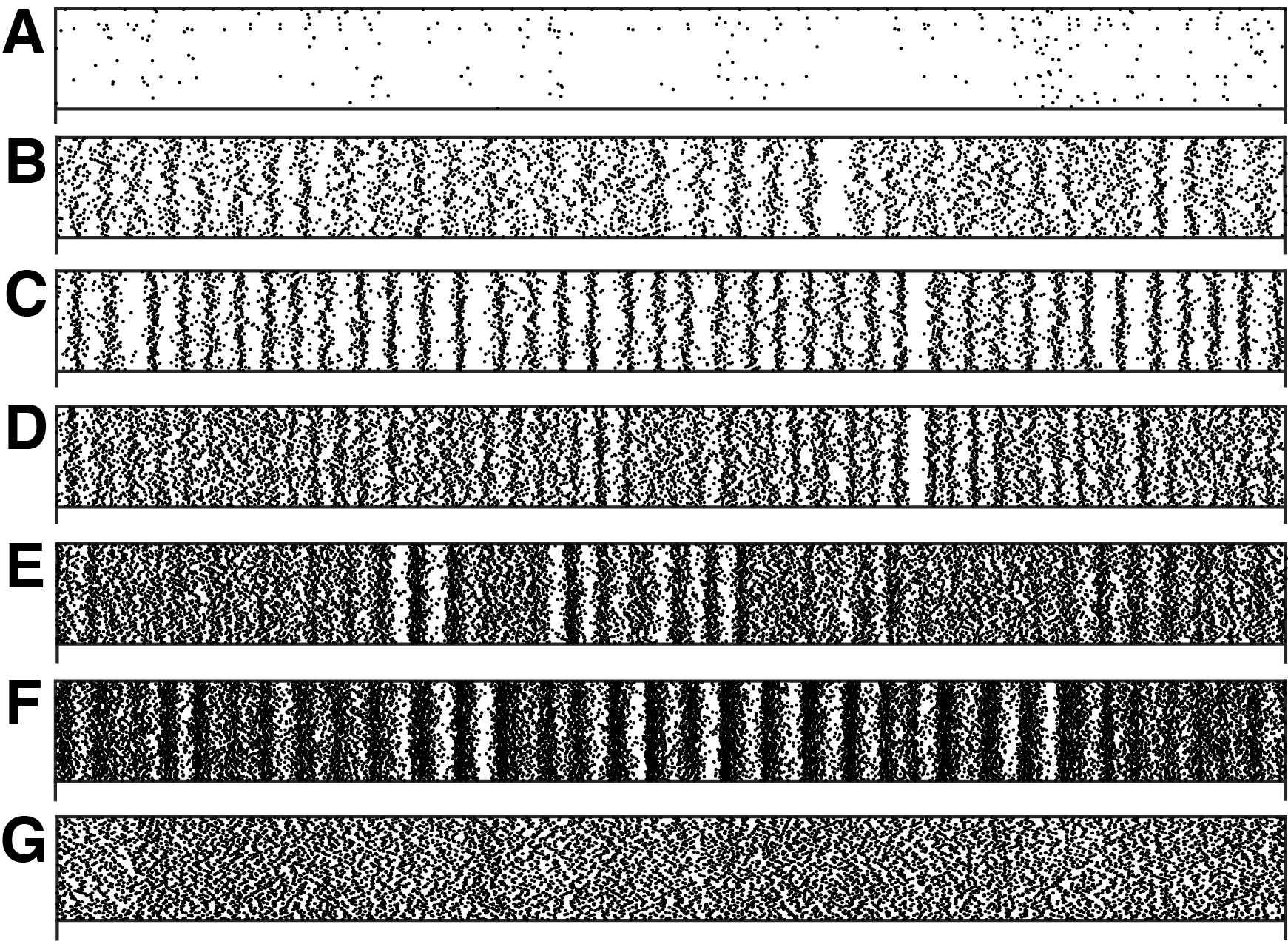}
\caption{Firing behavior at different overall coupling strengths $W=0.14, 0.18, 0.24, 0.34, 0.44, 0.54, 0.9$, from top to bottom. Rasterplots; vertical axis: neuron number, horizontal axis: time (A-F: 10000, G: 2000 time steps).}
\label{overview}
\end{figure}

In the following we analyze the change in firing of the characteristic networks, as $W$ progresses over the $[0,1]$-interval. Results pertaining to the location and the characterization of the avalanche critical point (see below) are based on a subset of at least $50$ simulations of distinct network realizations tied to the centers of the parameter distributions (Figs. 7 and 8); results of regarding the full characterization across the whole range of $W$ are from 10 chosen networks (Fig. 15). Lyapunov exponent calculations were performed by considering the network as one big dynamical system \cite{Kanderschaos} (cf. Appendix). Avalanches were extracted using the method pioneered in Ref. \cite{Beggs2003}, using time bins of size $\Delta_t$ equal to the average inter-event interval. An overview of the observed firing behaviors across the $W$-interval is provided in Fig. \ref{overview}. 

More specifically, we characterize the network state obtained at interaction strength $W \in [0,1]$ in Figs. 4-6 and 9-14, from left to right, top to bottom, as follows:
{\bf First panel}: Firing pattern, where the first row corresponds to the ``leader neuron''.
{\bf Second panel}:  Largest Lyapunov exponent (red dot), embedded in its dependence when varying $W$ (grey dots). For simpler biological interpretation, the obtained values are given in measures of $s^{-1}$, by identifying one Rulkov iteration with a duration of $0.5 \,ms$ \cite{Rulkov2004};  
from this, conventional, dimensionless, values can easily be recovered by the reader.
{\bf Third panel}: Kolmogorov-Sinai entropy (red dot, sum of positive Lyapunov exponents), embedded into the dependence obtained from varying $W$ (grey dots).
{\bf Fourth panel}: Avalanche size distributions (log-log scale).
{\bf Fifth panel}: Avalanche lifetime distributions (log-log scale).
{\bf Sixth panel}: First 128 exponents of the Lyapunov spectrum (cf. Appendix), where a green dot exhibits the largest exponent.

\begin{figure}[!!!!!!!!!!!!!!!!httttttttttttttttttttttttttttttttt!!!!!!!!!!!!!]
\includegraphics[width=0.75\linewidth]{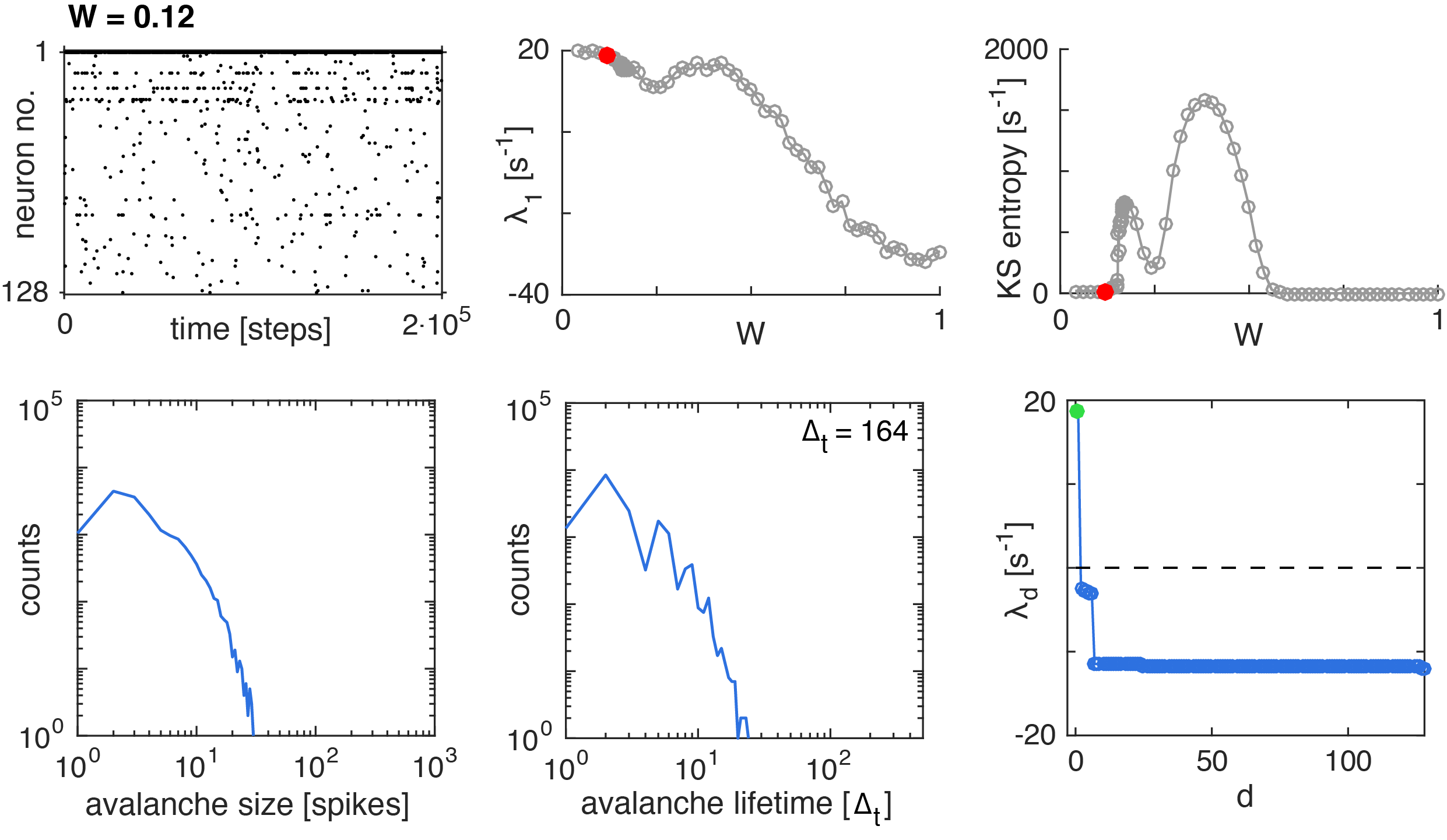}
\caption{(For a description of the displayed panels, see text). Network at $W=0.12$ at increased firing activity. Up to avalanche criticality, overall firing activity remains at an almost unchanged level. Lifetime-distribution reflects of the presence of a leader neuron.}
\label{Plot1}
\end{figure}

\begin{figure}[!!!!!!!!!!!!!!!h!!!!!!!!!!!!!!!!!!!!!!!!!!!]
\includegraphics[width=0.75\linewidth]{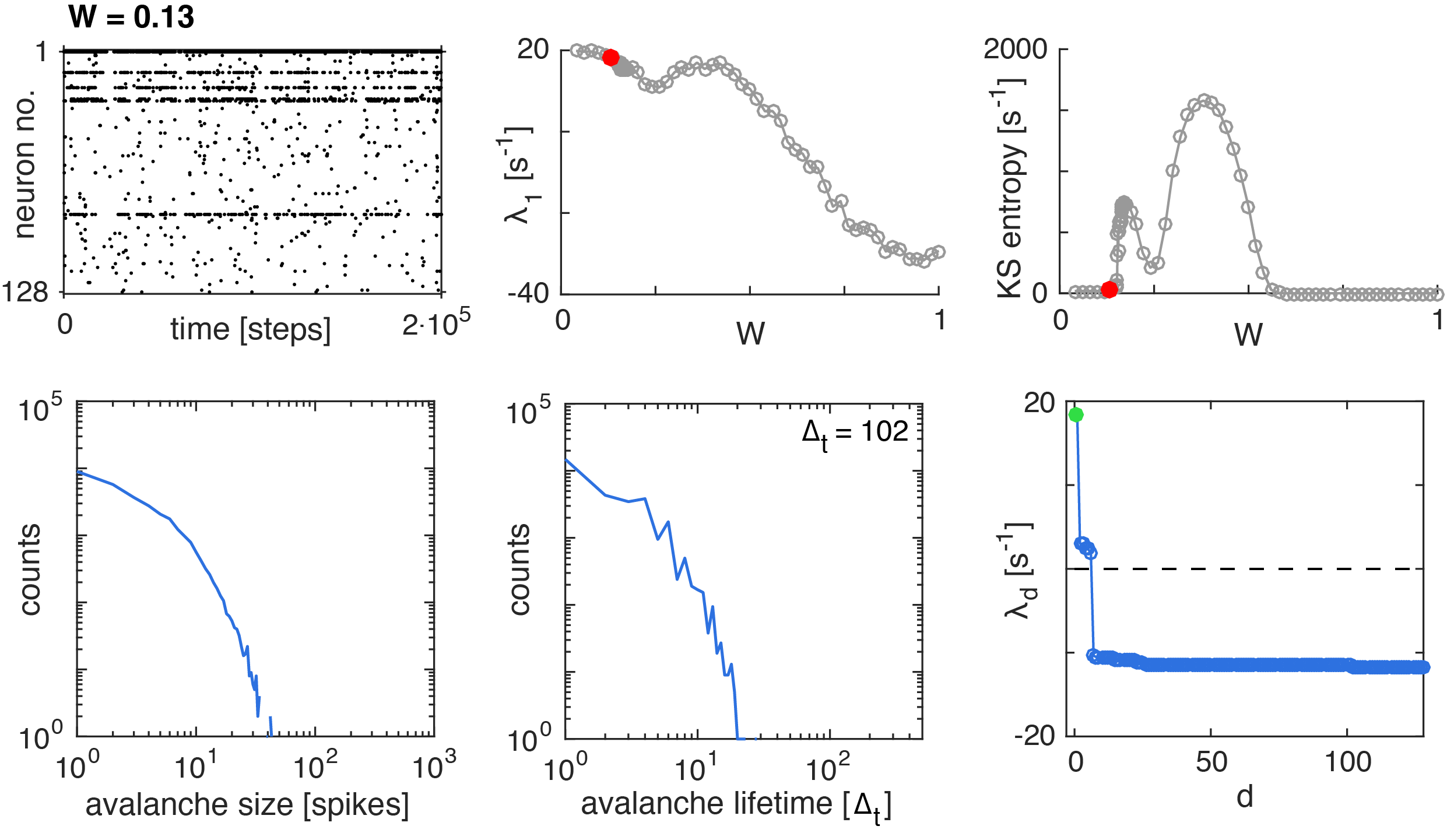}
\caption{Towards avalanche criticality: At $W=0.13$, additional modes become unstable.}
\label{Plot2}
\end{figure}

In our series of figures, we provide a full account on all prominent changes that the network behavior exhibits during a change of $W$, where our emphasis is on the Lyapunov spectra and the avalanche distribution properties. 
Starting our exploration at $W=0.1$, the network exhibits one positive Lyapunov exponent. Increasing the coupling steadily  leads, around $W=0.12$, to the separation of 3-4 neurons of reduced stability, from the bulk of negative exponents (Fig. \ref{Plot1}). 
At $W=0.13$, the additional nodes become unstable, where avalanche size and avalanche duration distributions remain to be of exponential type (Fig. \ref{Plot2}). 

\section{Avalanche critical point}

Around $W=0.14$, substructures emerge among the still negative exponents \cite{Kanderschaos}. More precisely, at avalanche criticality $W=0.139$, we observe an avalanche second order critical point. Its existence is verified 1) by statistically significant power-law distributions \cite{Deluca2013}, 2) by the existence of a scaling function \cite{Friedman2012}, and 3) by the fact that the crackling noise relation among the critical exponents is satisfied \cite{Sethna2001}. For more details see \cite{Kanderschaos}. At this point, the firing rate increases and the largest Lyapunov exponent decreases, whereas the number of positive exponents dramatically increases. 

\begin{figure}[!!!!!!!!!h!!!!!!!!!!!]
\includegraphics[width=0.75\linewidth]{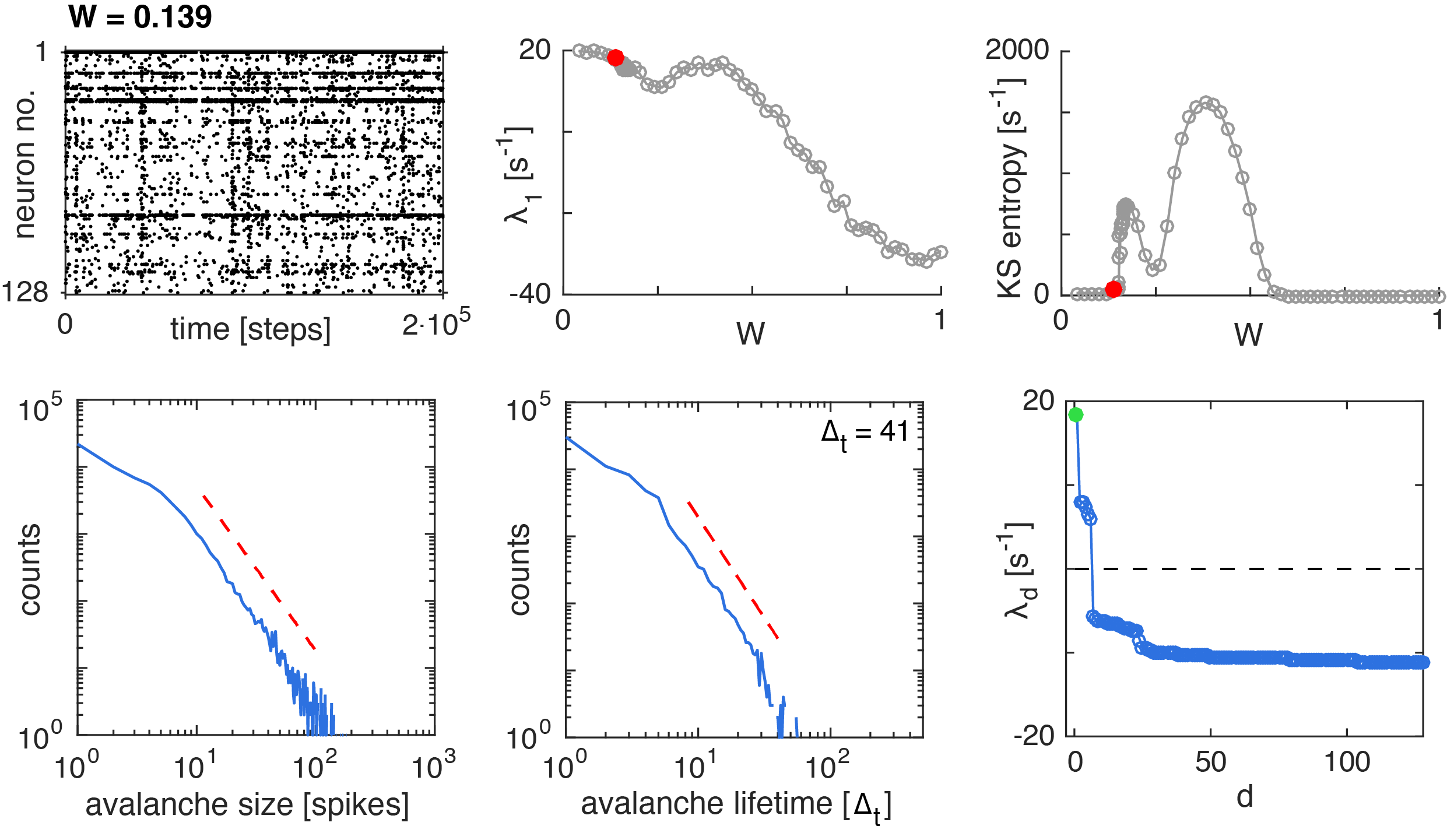}
\caption{At avalanche criticality ($W=0.139$), an avalanche size power-law decay exponent of $\alpha \approx 2.45$ is obtained (using the method of Ref. \cite{Deluca2013}).}
\label{Plot3}
\end{figure}

In the subcritical case ($W<0.139$), the avalanches are generally small, where their size distribution decays exponentially. In the supercritical case ($W>0.139$), we observe an increased number of large avalanches resulting in a characteristic hump toward the end of the distribution \cite{Lorimer}. A similar metamorphosis of the distribution shape is observed for avalanche lifetimes. At critical avalanche behavior, the lifetime distribution also follows a power law  (Fig.~\ref{Plot3}, exponent $\tau \approx 3.0$), of a somewhat lesser quality if compared to the size distribution, a commonly observed phenomenon in electrophysiological experiments \cite{Beggs2003,Pasquale2008, Hahn2010}.

\begin{figure}[!!!!h!!!!!!]
\includegraphics[width=0.75\linewidth]{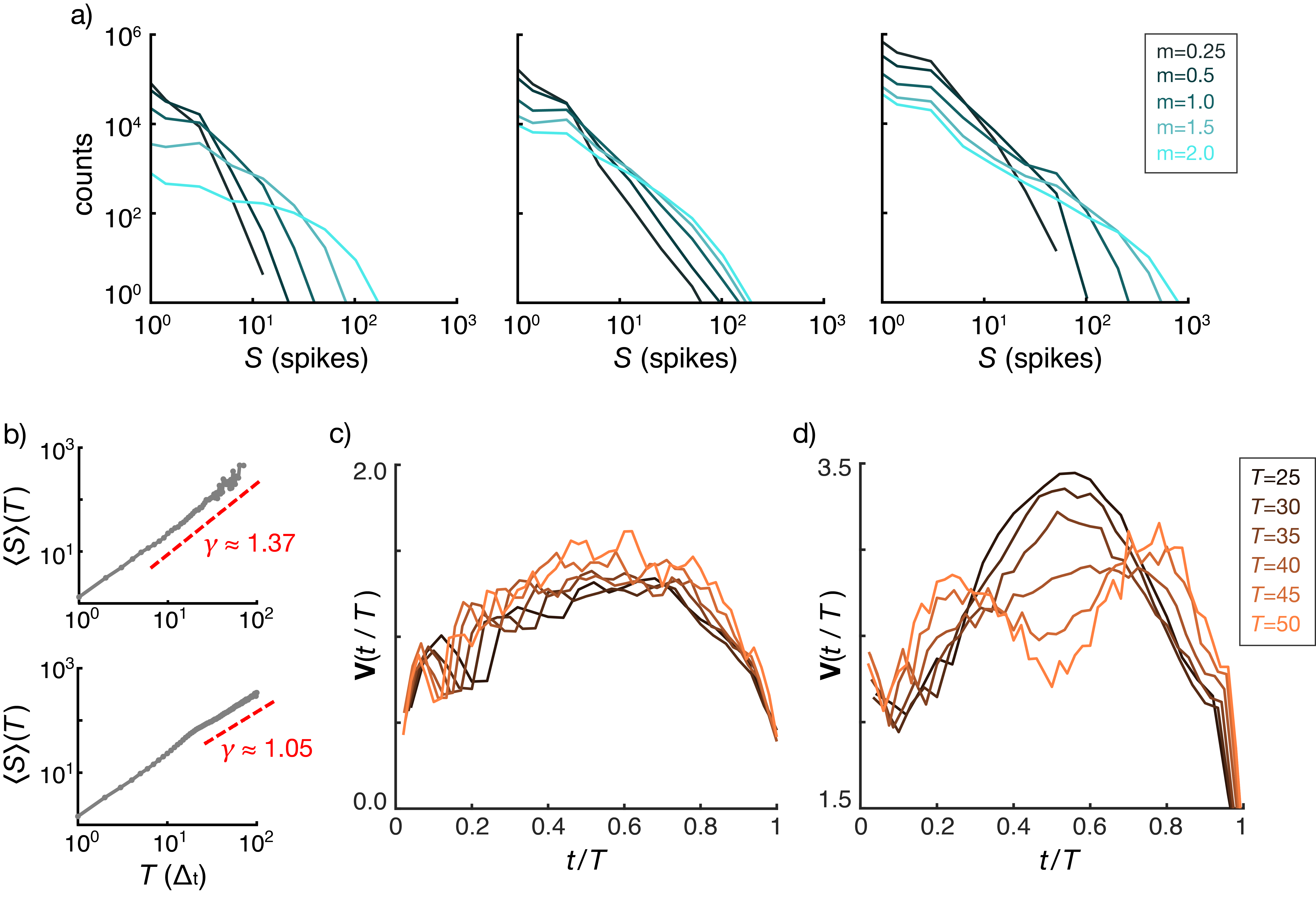}
\caption{Tests of criticality: a) Avalanche size distributions based on time bins of size $\tilde{\Delta}_t=m\cdot \langle IEI \rangle, \,\, m\in \{0.25,0.5,1.0,1.5,2.0\}$, for subcritical (left, $W=0.13$), critical (middle, $W=0.139$), and supercritical states (right, $W=0.15$), where logarithmic histogram binning was used. b) Mean avalanche size as a function of lifetime, for critical (top) and supercritical (bottom) states. The red dashed line exhibits a power law relationship $\langle S \rangle(T) \propto T^{\gamma}$. c)~Avalanche shapes of the critical state show a noisy collapse ($T \in\{ 25,30,\ldots,50\}$, from darker to lighter color), expressing a high degree of self-similarity, but also the particular role of the intrinsically firing neuron (see \cite{Kanderschaos} for details;  the collapse is substantially improved if the role of the leader neuron is randomly distributed to neurons of the network). d) For the supercritical state, rescaling does not lead to a collapse ($T \in \{25,30,\ldots,50\}$).}
\label{tests}
\end{figure}

\begin{figure}[h]
\includegraphics[width=0.3\linewidth]{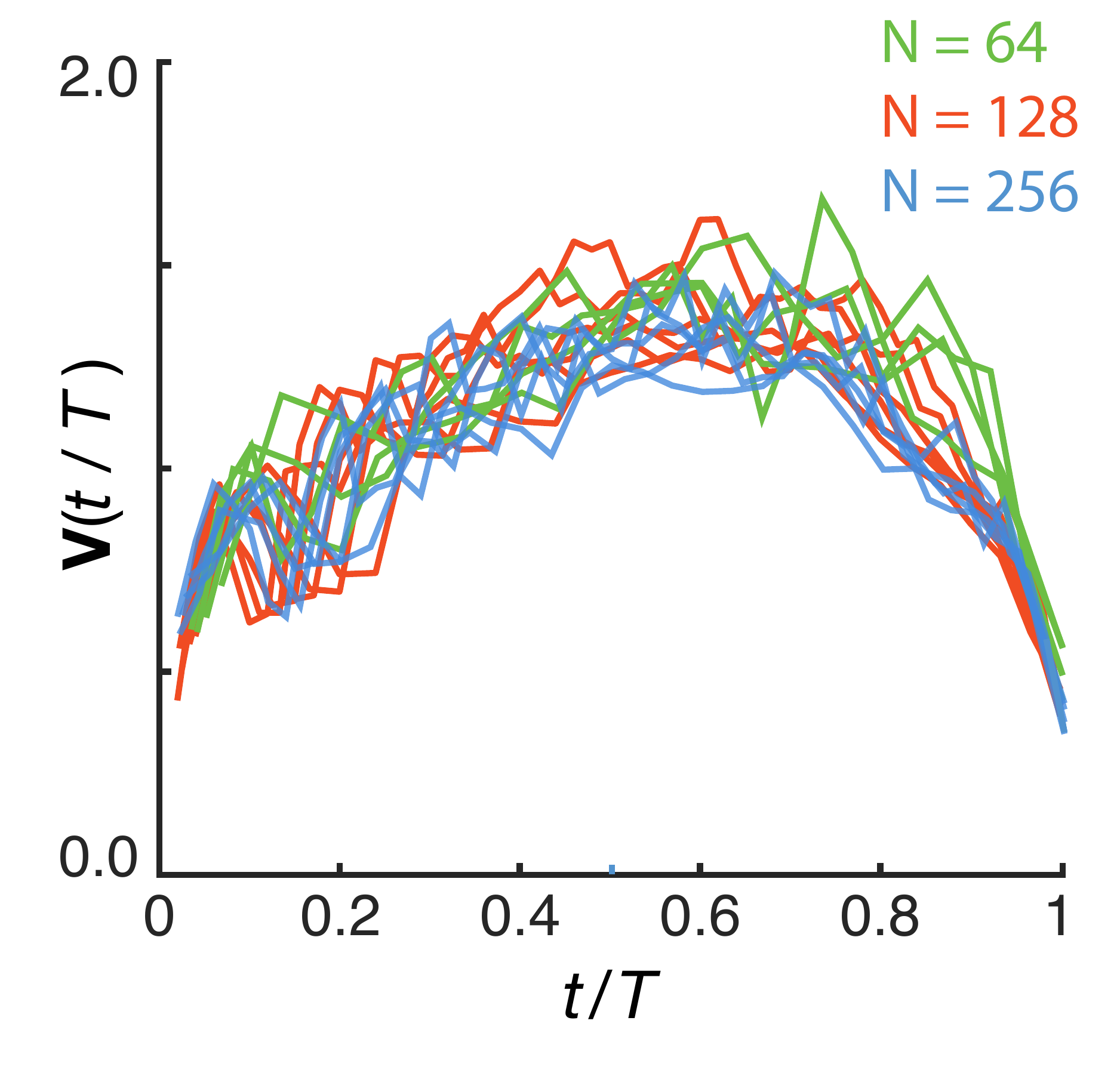}
\caption{Scaling in system size $L$ across the columnar scale, with $N$ denoting the number of neurons in the network. Our results for $N \in \{64,128,256\}$ collapse. Beyond this range, strongly outside of the columnar scale, an appropriate scaling of the connectivity (and of the number of nucleation sites) would need to be provided, which is nontrivial (see our Introduction section).}
\label{tests2}
\end{figure}

Power law distributions can be caused by several mechanisms and are only necessary, but not sufficient for criticality. Several tests were performed to ensure that we are at an avalanche critical point indeed. For genuine scale-free behavior, the choice of the temporal bin size, $\tilde{\Delta}_t$, should not affect the avalanche size distribution (cf. \cite{Priesemann2014}). Fig.~\ref{tests}a) exhibits that in the critical case, choosing different binnings $\widetilde{\Delta}_t= m \Delta_t, \, \,m\in \{0.25,0.5,1.0,1.5,2.0\}$ only mildly affects the distribution; in the subcritical and the supercritical cases,  the effects are markedly stronger. We, moreover, checked whether all avalanches of the critical state would collapse to one characteristic shape, upon a corresponding rescaling of time \cite{Sethna2001, Friedman2012}. To this end, the shape $V(T,t)$ of an avalanche of lifetime $T$ is calculated. $V(T,t)$ expresses the temporal evolution (time variable $t$) of its 'shape' measured by the number of spikes emitted in the temporal bin around time $t$. For each $T$, the average avalanche shape  $\langle V \rangle(T,t/T)$ is calculated.
From the scaling ansatz between the mean size of avalanches $\langle S \rangle(T)$ and their lifetime $T$,
$\langle S \rangle(T) \propto T^{\gamma}$, the critical exponent $\gamma$ is obtained.
From this, the universal scaling function representing the characteristic shape of all avalanches, emerges as $\textbf{V}(t/T) = T^{1-\gamma} \langle V \rangle(T,t/T)$. 
For our 'critical' network, $\langle S \rangle(T)$ indeed follows a power law, with exponent $\gamma \approx 1.37$ (Fig.~\ref{tests} b)). In the case of our supercritical network, a smaller range of the function also follows a power law, which permits the comparison between the self-similarities of the avalanche shapes of the critical and the supercritical state. For the critical network, we observe a noisy collapse of the avalanche shapes of duration $T \geq 25$ (Fig. \ref{tests}c)). For shorter lifetimes ($T < 25$), the avalanche shapes fail to exhibit a nice collapse because of the strong influence of the intrinsically firing, non-scaling, neuron. If, however, the role of the leader neuron is randomly distributed among the neurons of the network, a much more precise collapse can be obtained, improved across the whole of the $t/T$ interval. Generally, universal scaling at shortest length scales cannot be expected to emerge reliably, as individual system part behavior can be stronger than the collective behavior at these scales \cite{Sethna2001}. In contrast, for the supercritical network, there is no such collapse (Fig. \ref{tests}d)).
As a final test, we examined whether the crackling noise relationship between critical exponents, $(\tau~-~1)/(\alpha~-~1)~=~\gamma$, would hold \cite{Sethna2001, Friedman2012}, for our critical network. The critical exponents of avalanche lifetime distribution ($\tau = 3.0$), avalanche size distribution ($\alpha = 2.45$), and the function of the mean avalanche size depending on the lifetime ($\gamma = 1.37$) fulfill the required relation remarkably well. Taken together, power law distributions, self-similarity of avalanche shapes, and an excellent fulfillment of the fundamental relation between critical exponents, strongly suggest that our 'critical' network indeed closely follows the critical network state template. 
Across the network sizes for which our model accounts, a nice collapse with respect to the system size is observed (Fig. \ref{tests2}). Beyond the typical size of a column that it is meant to describe, our network architecture template is non-scaling. For larger network size, modifications in the network structure (number of leader neurons, the network's wiring, etc.) would be needed. Comparing mouse and humans demonstrates that the number of synapses are likely to remain and also the number of myelinated fibres, e.g., seems not to scale. From our view, a full scaling theory of the mammalian cortex is therefore unlikely by close-to-biology computational models.

Important to note is that avalanche criticality happens markedly away from dynamical `edge of chaos' criticality that would request a zero leading network Lyapunov exponent. While the positivity of this exponent appears evident from construction, it must be nonetheless be verified, as the coupling to non-chaotic nodes could in principle suppress the chaotic driving node dynamics.
The inspection of the network Lyapunov exponents reveals a change from one leading and immediately affected unstable direct neighbors around avalanche criticality, to many more unstable nodes at supercriticality. During this transition, the firing rate of the driving node decreases, but synchrony increases, as the driving node drags the other nodes increasingly along. This leads to a first peak in synchronization where roughly half of the nodes take part in this process. This interpretation is corroborated by the hierarchical structure of the spectrum of Lyapunov exponents. This process ends in a first collective mode. The measured avalanche size power-law exponent of $\alpha = 2.45$ exhibited by our biologically plausible network is similar to $\alpha=2.5$ observed in simulations of bursting recurrent networks, for 'background' avalanches \cite{Jordi2013} (that have been linked to critical percolation on a Cayley tree-like network \cite{Stauffer1994}).
Compared to experimental settings that have yielded avalanche size distribution exponents $\alpha \in (1.5, 2.6)$  \cite{Mazzoni2007,Tetzlaff2010,Shew2015}, and compared to results from the analysis of local field potential avalanches \cite{Beggs2003} and to cochlear activation networks \cite{stoopprl2016}, for which exponents $\alpha \simeq 1.5$ have been obtained, our exponents appear somewhat augmented. However, during the maturation of biological neuronal networks, a large variety of different states with largely varying characteristics have been found as well \cite{Berger2015}.

Even though strong qualitative changes on the topological network state were enforced across an extremely large parameter range, our network maintains its chaotic dynamics. This demonstrates that avalanche criticality does not necessarily co-occur with edge of chaos criticality. Rather, this suggests that in neural networks with non-trivial node dynamics, two separate phase transitions may occur. 
The high variability of exponent $\alpha$ may express the different network and dynamical conditions under which avalanche criticality is possible, and may point to a link between avalanche and edge of chaos criticality, albeit of a much weaker form than is usually assumed (generally, we expect that higher values of this exponent will be related to stronger computational performance). For completeness, we note that the computational significance of avalanche criticality has been discussed in Ref. \cite{stoopprl2016}.

\section{Multiple transitions beyond avalanche criticality}

After passing the critical avalanche point, the firing frequency of the intrinsically firing neuron, but even more so of the other neurons of the network,  increases abruptly. A decrease of the largest Lyapunov exponent, accompanied by an increase of the Kolmogorov-Sinai entropy, suggests that individual neurons coordinate their activities by synchronization (which is corroborated by the calculated synchronization measure, Fig. 15). Further increase of $W$ now leads to a consistent increase in firing frequency.
At $W=0.15$ (Fig. 9), large-scale temporal firing patterns become apparent also in the raster plots. Roughly half of the first 128 Lyapunov exponents have become positive at this point, whereas the distributions of the avalanche size and avalanche time duration are still vaguely power-law reminiscent. This continues up to  $W=0.169$ (Fig. 10), where a maximal Kolmogorov-Sinai entropy and the fact that all first 128 Lyapunov exponents are positive indicate that a maximal state of chaoticity has been reached. The observed paradigm suggests that the immediate reaction upon the critical avalanche transition is softened, because a distribution onto a large number of degrees of freedom in the network is possible. After this first state of maximal chaoticity, a substantial increase in $W$ is required to bind the first Lyapunov exponent to the bulk of the exponents, leading to a second, global, state of maximal chaoticity, until further increased coupling destroys the coherency of the network dynamics required for positive Lyapunov exponents.

\begin{figure}[hhhhhhh!!!!!!!!!!!!!!!!]
\includegraphics[width=0.85\linewidth]{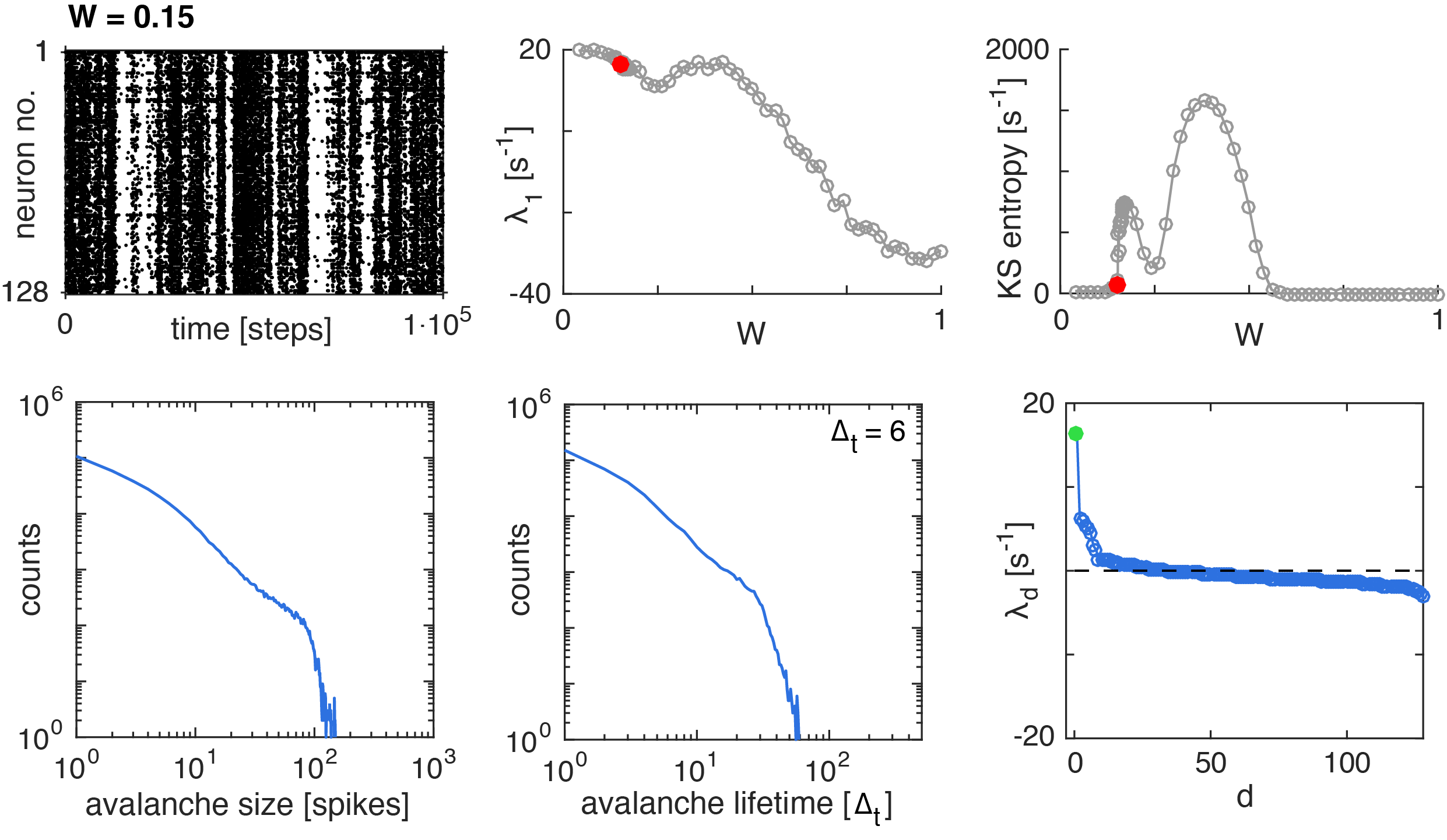}
\caption{After avalanche criticality, we observe a rather abrupt increase in firing activity, as could be expected from an order parameter in the vicinity of a critical state. The strong increase continues up to $W=0.15$; at higher values $0.15<W<0.6$, the increase with $W$ becomes more moderate.}
\label{Plot4}
\end{figure}
\begin{figure}[hhhhhhhhhhhh!!!!!!!!!!!!!!!!!!!!!!!!!!!]
\includegraphics[width=0.85\linewidth]{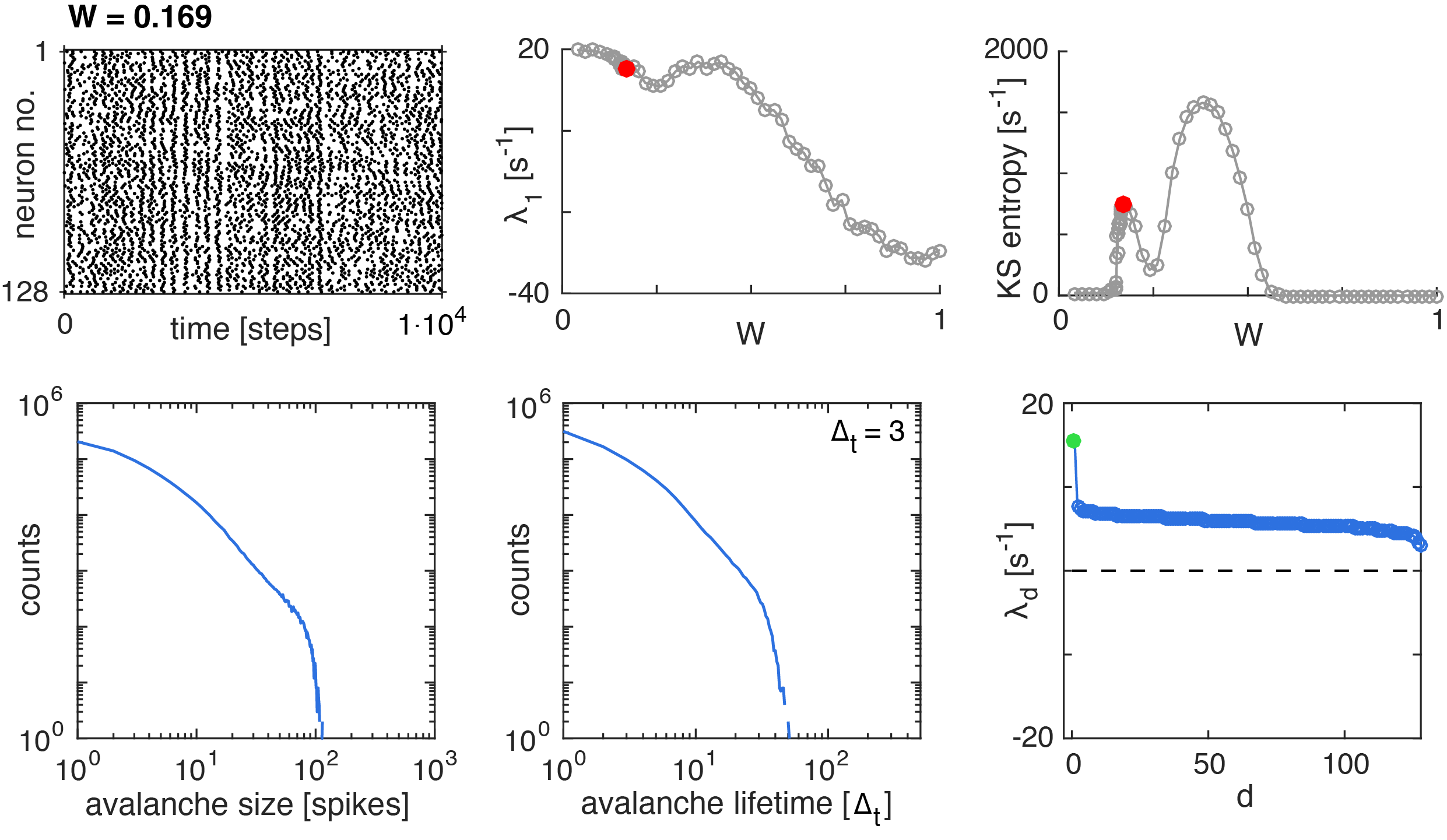}
\caption{$W=0.169$: First local maximum of the Kolmogorov-Sinai entropy, based on temporally extended firing patterns. Although the distributions have a faint power-law appearance, the data is not invariant to binning. }
\label{Plot6}
\end{figure}
Upon a further increase of $W$, the characteristic hump indicating supercritical behavior finally becomes apparent at $W=0.22$ (Fig. 11). At this point, an excess of $W$ beyond maximal exhaustion of the individual degrees of freedom starts to bind the individual activities. A large synchronization measure (Fig. \ref{overview2}) suggests that the negative exponents are re-established by means of synchronized network activity.  System-sized firing patterns now dominate the raster plots, the largest Lyapunov exponent decreases and some of the first 128 exponents become negative. 
\begin{figure}[hhhhhhh!!!!!!!!!!!!!!]
\includegraphics[width=0.85\linewidth]{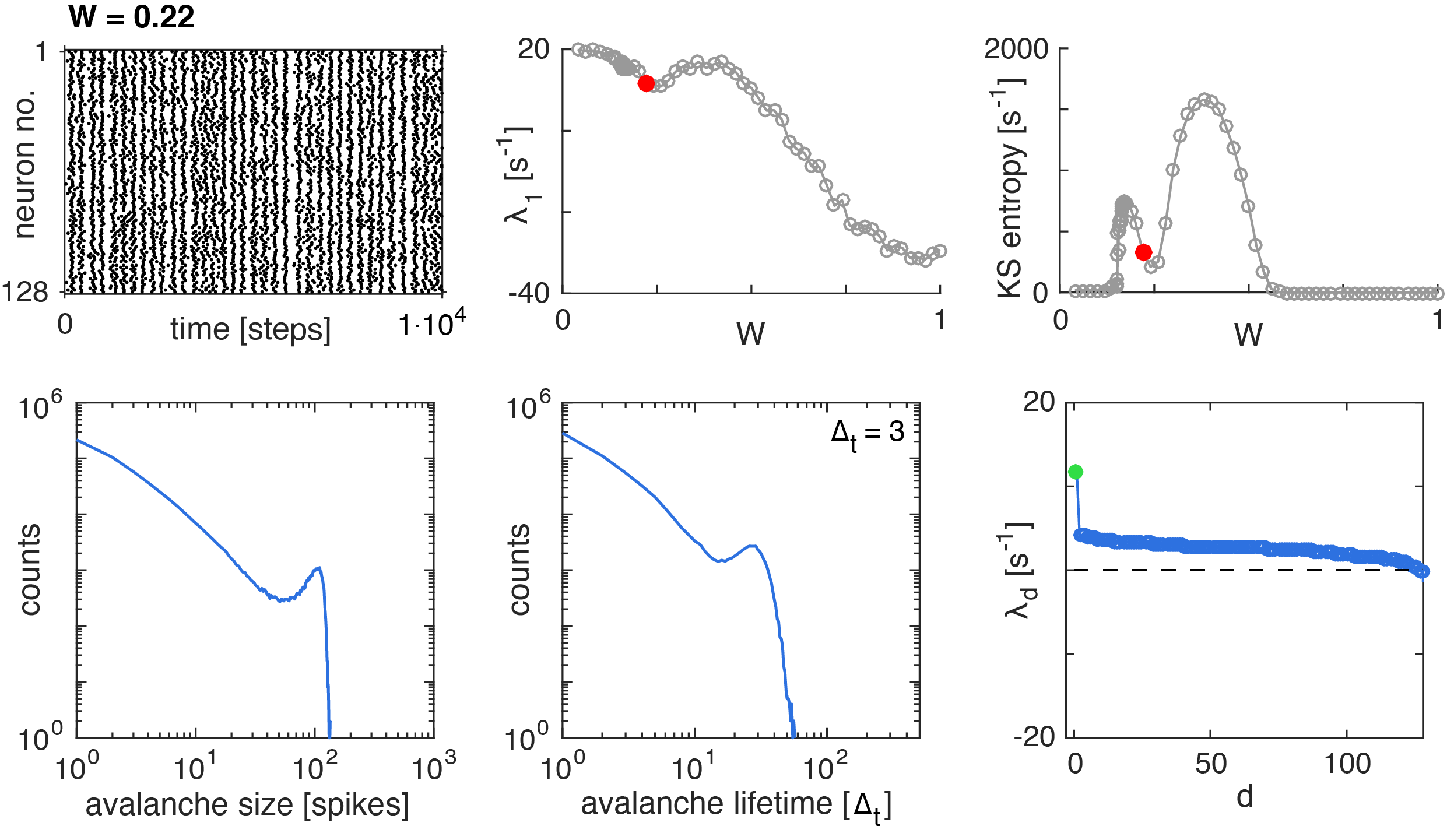}
\caption{Emergence of preferred patterns at $W=0.22$, visible in the avalanche distributions.}
\label{Plot7}
\end{figure}
\begin{figure}[hhhhhhhhhhhhhh!!!!!!!!!!!!!!!!!!!!!]
\includegraphics[width=0.85\linewidth]{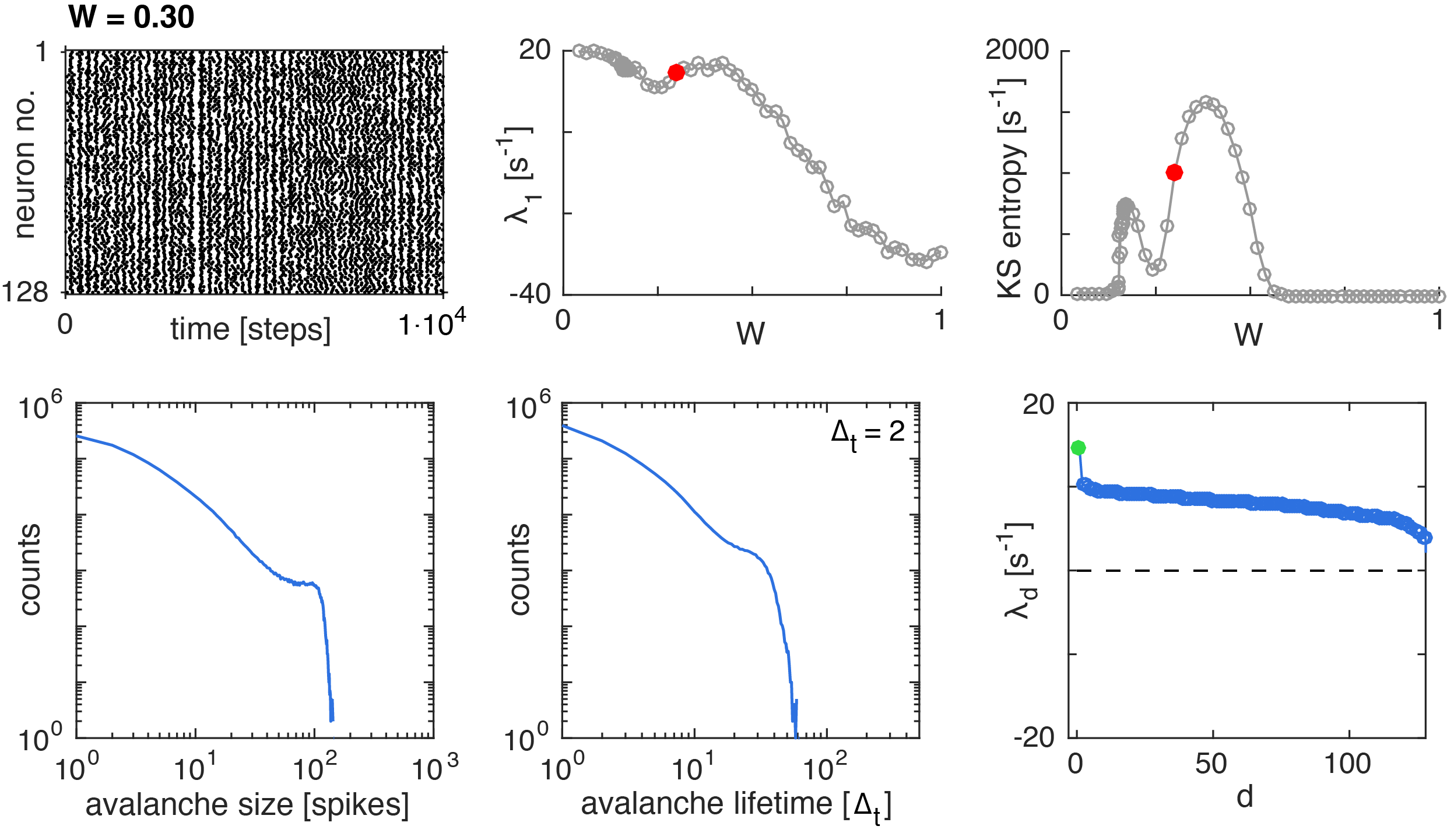}
\caption{ $W=0.3$: Increase of the largest exponent again; following exponents are dragged to more positive values. The first exponent still leads the rest of the exponents. Beginning of the fully developed network firing phase.}
\label{Plot8}
\end{figure}
At $W=0.3$ (Fig. 12), the largest exponent is again increasing, and additional exponents are dragged to more positive values as well. 
This process is completed at $W=0.38$ (Fig. 13), where local maxima of both individual positive Lyapunov exponents and Kolmogorov-Sinai entropy express that a maximal degree of chaoticity has been reached. At this point, the dominance of firing patterns of system size fades, giving way . 
\begin{figure}[]
\includegraphics[width=0.85\linewidth]{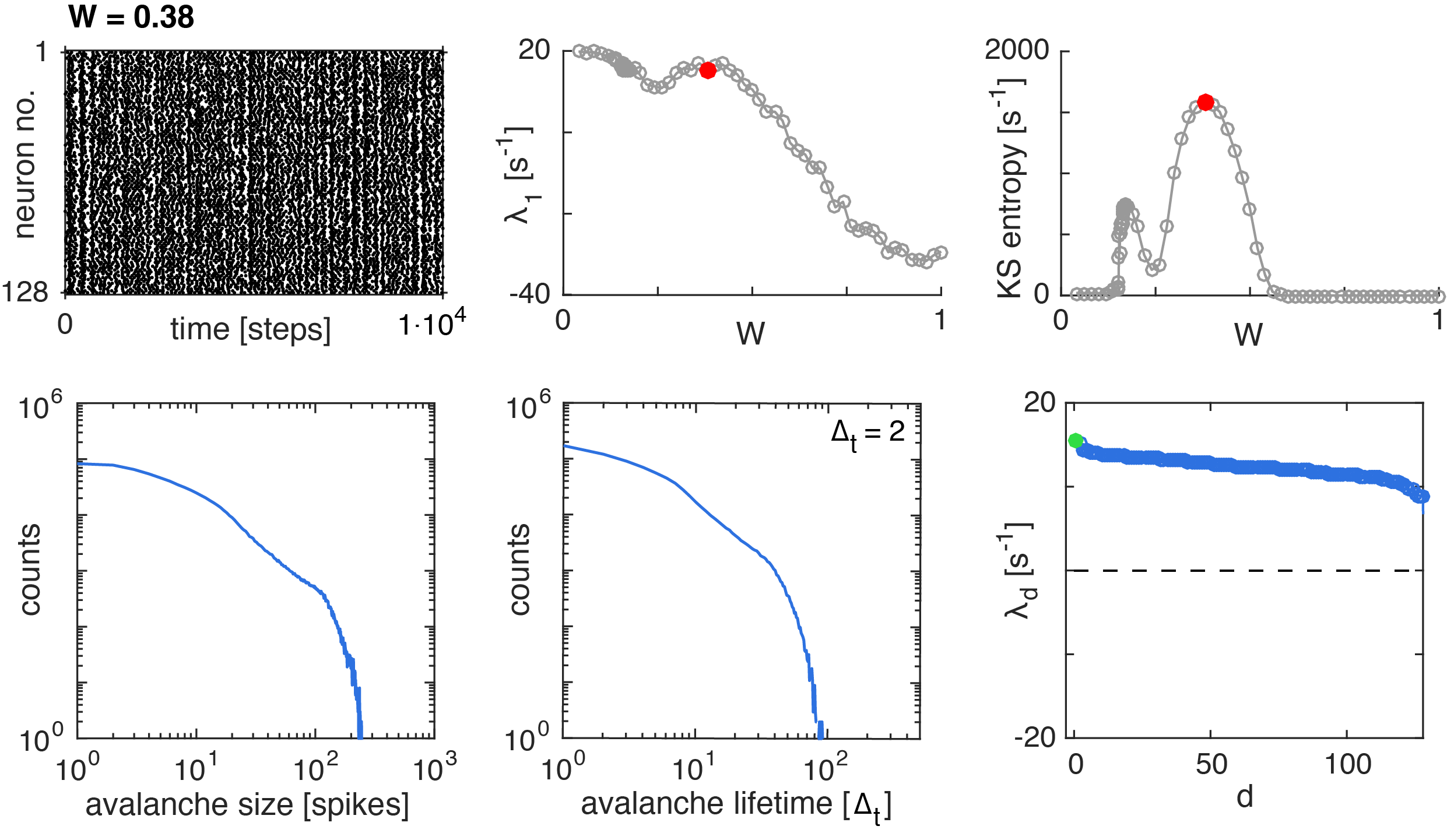}
\caption{$W=0.38$: Second local maximum of the Kolmogorov-Sinai entropy, based on a temporally extended firing pattern phase. The first Lyapunov exponent has fully merged with the bulk of the remaining positive exponents. End of the fully developed network firing phase.}
\label{Plot9}
\end{figure}

Thereupon, controlled by synchronization, a decrease of all exponents sets in. At $W=0.6$ (Fig. 14), this process is completed, with maximized synchronization, but no positive Lyapunov exponent remaining. The transition point from positive to negative leading exponent (at $W=0.59$), has some resemblance with edge-of-chaos criticality. The detailed avalanche analysis performed with the standard tools, failed, however, to provide convincing evidence for this, and rules out a case of coincidence of avalanche with edge-of-chaos coincidence at a high level of confidence.

\begin{figure}[!!!!!!!!!!!!!!!!!h!!!!!!!!!!!!!!!!!!!!!!!!!!!]
\includegraphics[width=0.85\linewidth]{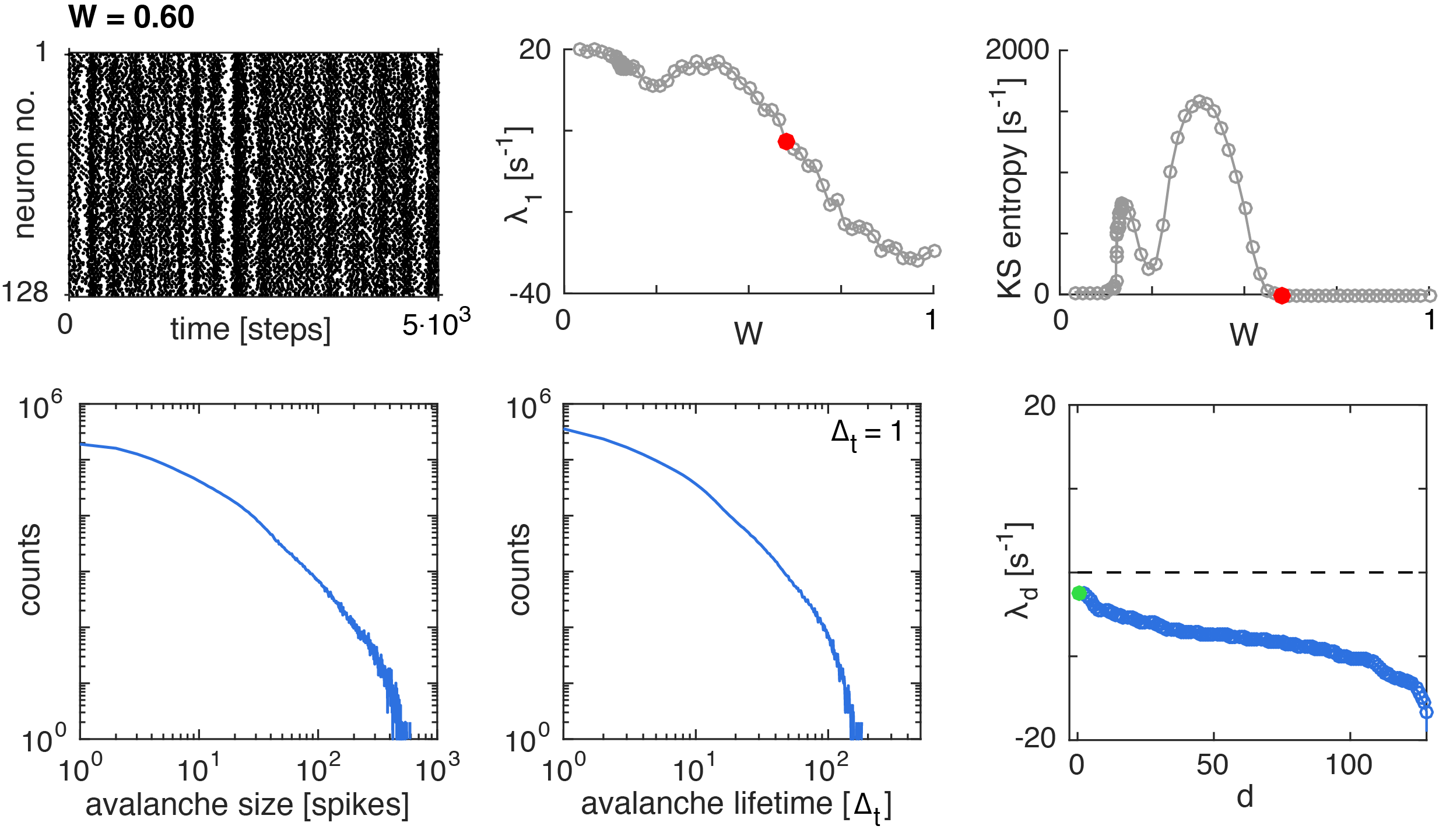}
\caption{$W=0.6$: No positive Lyapunov exponents survive at a local maximum of synchronization. At the transition point  from positive to negative leading Lyapunov exponent, the avalanche distribution is very similar to the one presented. Although a small linear slope part is exhibited, the distribution properties rule out avalanche criticality with high confidence.}
\label{Plot11}
\end{figure}
\begin{figure}[!!!!!!!!!!!!!!!!!!!!!!!!!!h!!!!!!!!!!!!!!!!!!!!!!!!!!!!!!!!!!!!!]
\includegraphics[width=0.75\linewidth]{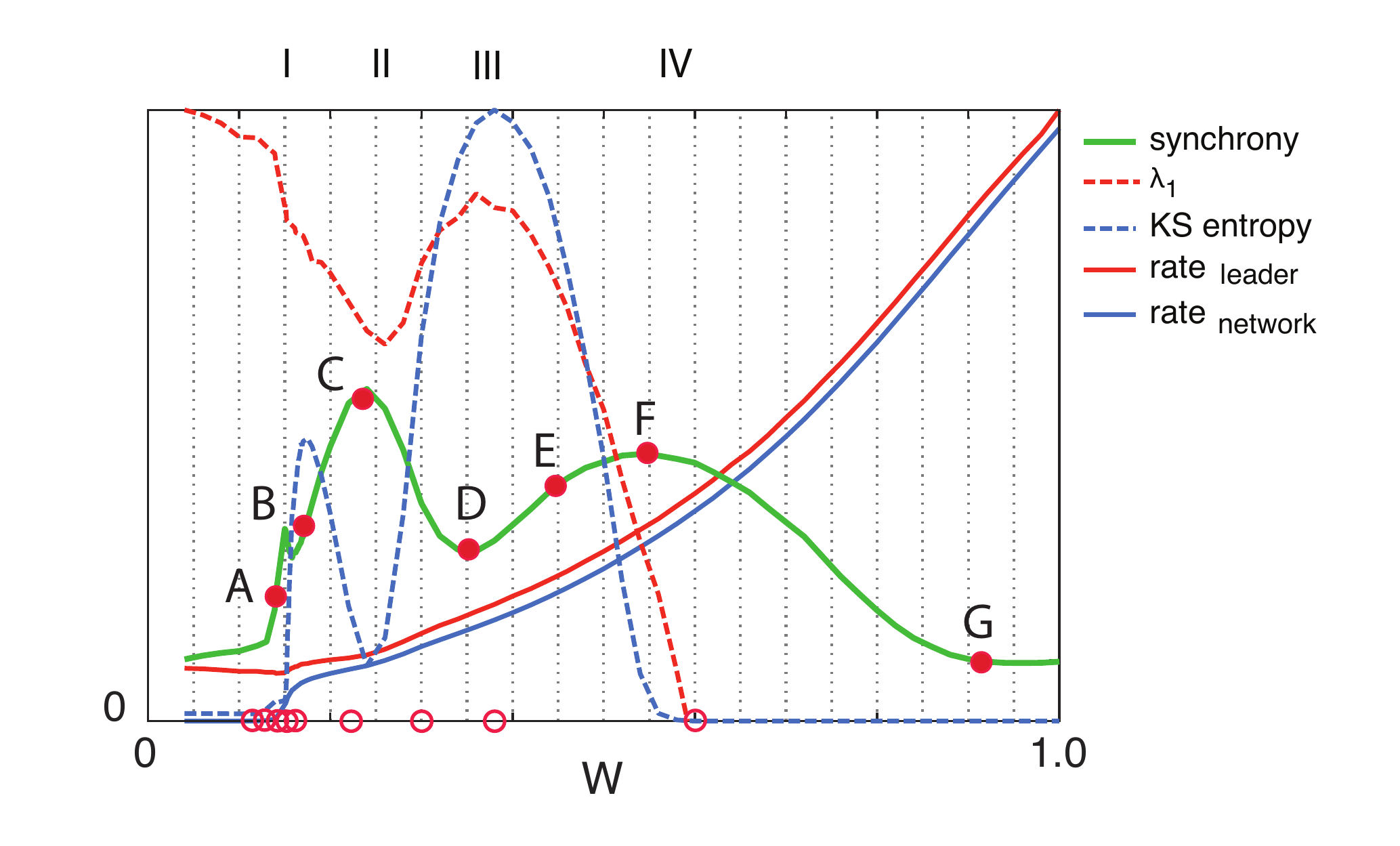}
\caption{Overview: Different network characterizations compiled (y-axis: arbitrary units). Avalanche criticality is characterized as the unique point where synchronization and KS-entropy increase dramatically, and where the leader neurons merge with the network. Green: Synchronization measure, based on the slow variables of neuron dynamics \cite{Rinzel}. Red dots: Parameter locations corresponding to the raster plots of Fig. \ref{overview}; open dots indicate the location of the $W$-parameters of the network states discussed in detail.}
\label{overview2}
\end{figure}
An overview on the various dynamical network transformations is given in Fig. \ref{overview2}, which are only poorly reflected in the firing rates.
After the avalanche critical point has been reached (I), we observe that the leader neuron becomes an essentially indistinguishable part of the network. This is indicated by a co-evolution of the largest Lyapunov exponent with the KS-entropy. The different network parts become largely independent. Their evolution is counterbalanced by synchrony that tames the Lyapunov exponents with larger coupling (II), by constraining independence. Further increased coupling then reduces synchrony (III), generating more numerous smaller independent subnets. Finally, even larger coupling removes their independence and tames the network's variability on stimulus as the basis for positive Lyapunov exponents.

\section{Conclusions} 

We demonstrated that over the whole range of biologically meaningful coupling, a coincidence of the two notions of criticality does not happen. This substantially expands the previous finding \cite{Kanderschaos}, where we provided a counter example to coincidence. The present work adds to the latter a notion of genericity, carefully worked out within the chosen model class. The scan that we performed across possible network behavior, using connectivity strength $W$ as the parameter, seems to be sufficiently general to underscore that the relationship between topological and dynamical criticality is not as simple as was previously thought. From our results, it emerges that at least two largely independent variants of criticality exist, either or both of which biological nature seems to make use of. The obtained results suggest that close-to biology networks must be expected to undergo whole  transformations of firing behavior (depending on input, physiological and biochemical parameters, measured as changes in physiological, dynamical, topological, or statistical physics observables). During these changes, we may find several situations where avalanche size or avalanche duration distributions appear to follow power laws. If in modeling studies, such cases can be identified as phase transition points (using finite size scaling analysis, as we have done here), they are of particular interest, as they should also be identifiable in biological experiments.  Conversely, identified biological phase transition points should be included as corner stones for statistical physics modeling of such networks. In this respect, our work provides important first explorative steps.

  In our mesoscale modeling, the influence of synchronization emerges on a similar time-scale as the change of the Lyapunov exponents (Fig. \ref{overview2}). Experimental evidence appear to support such a situation more than models of balanced networks, where synchronization plays a much less prominent role. Our results underscore a largely antagonistic role that synchronization and dynamical exponential separation have, beyond the avalanche critical point. In principle, from the basic theory of dynamical systems \cite{Peinke1992}, such clear roles  cannot be expected, as coupling can enhance instability as well as stability; what it does at a given parameter largely depends on the coupled system's topology which is difficult to get a grip on for inhomogeneous networks that we deal with.  We expect that the behavior of the present case is due to the large number of degrees of freedom that we deal with.
 The independence and complementarity of the used network characterizations (Fig. \ref{overview2}) exhibits that many structural transformations in network dynamics shape and reshape the network behavior. Transformations that may look obvious in terms of particular observables, may not be easily visible in other observables, in this way expressing the challenge our understanding of neuronal dynamics is confronted with as soon as we go beyond strongly simplified models. In function of the free parameter $W$, these measures indicate the presence of at least four distinct transitions that we expect to be generic for this network type.  The transition points in the space of primary dynamical characteristics of collective behavior ('characteristic space') classify the structural changes in firing behavior much more precisely than is possible with conventional methods (e.g. raster plots, isolated measures).  Following the structural changes in firing behavior on increasing connection strength in this space, leads to a much needed improved quantitative understanding of these processes. For real-world networks, the extraction of the 'characteristic space' may be cumbersome, unreliable, or even be impossible for some networks, and the presence and succession of the four transitions may even vary, as a uniform increase of the networks connection strength, as required by parameter $W$, may not be realistic in these networks. We believe, however, that the observed succession of transitions provides an important template for understanding generic changes underlying variable network firing behavior.
 
 While we do not claim that the whole of the modeled $W$-interval needs to make biological sense, we  point out that the mesoscopic synapses encompass the bundled effect of all the microscopic synapses impinging from one neuron onto a target neuron, expressing, in particular, synchronization effects of the microscopic level. Seen from the average firing rate of a single neuron in our network, the observed  $\sim 4 \, spikes/1000$ steps at $W = 0.2$, and $\sim 16 \, spikes/1000$ steps at $W = 0.6$ equate to about 8 and 32 Hz, respectively, if one time step is equated to 0.5 msec (Rulkov's recommendation \cite{Rulkov2004}). Such values are often observed in the biological example (e.g., from neurons of the cat visual cortex V1 \cite{Baddeley}) and should therefore not be seen as unrealistic. In the context of MEA's, firing rates have been found to strongly depend on the composition of the media. Varying these media can lead to an increase of the firing rate by even more than of a factor of four (preliminary data from enriched media cultures, Yoonkey Nam Lab, KAIST, Korea). Enriched media cultures often show distinct firing behaviors compared to normal cultures, where the origin of the effect has not been sufficiently clarified.

To what degree the observed paradigm of network evolution, in dependence of neural connectivity strength, might be of a general nature (or even be `universal' for a larger class) needs to be explored by more extended modeling studies and by comparing with biological measurements. 
A generic nature of our observations, however, is highlighted by networks without leader neurons, where we observe in all observables the dependence exhibited in Fig. \ref{overview2}, after the avalanche critical point. This indicates that the transitions (I-IV) are fully described in terms of the Kolmogorov-Sinai entropy. In case where we have no distinguished leader neuron, generalized network states take the role of the associated eigenvectors (they largely depend on the individual network characteristics for small $W$) ; the process of binding the largest exponent to the bulk network then transforms into enforcing toward an optimal distribution of generalized network states that depend to an ever lesser degree on the network's original configuration, by means of increasing $W$.
Moreover, we have indication that the paradigm in its general dependence on a rescaled $W$ (existence of a critical avalanche state without co-incidence with edge-of-chaos, power-law exponent $\alpha \sim 2.45$) is not abandoned if Rulkov node dynamics is replaced by integrate and fire dynamics (see Appendix). Finally, we observed a very similar behavior in a recently analyzed model of an elementary adaptive network automaton \cite{Wechsler}.
This all hints at a very general nature of the observed network transformation paradigm.

Our approach of the detailed, close-to-biology modeling seems a rather promising one, since in experiments on neural tissue, the strength of neuronal coupling can be manipulated across a similar range compared to our modeling study. From this, we expect guidance for further modeling and insight into the nature of the corresponding biological examples. 



\section*{Funding}
This work was supported by the Swiss National Science Foundation grant (200021 153542/1), an internal grant of ETHZ (ETH-37 152), and a Swiss-Korea collaboration grant (IZKS2 162190) (all to RS) and a JSPS Grant-in-Aid for Challenging Exploratory Research 26540127 to YU.

\section*{References}

\bibliographystyle{frontiersinHLTH&FPHY} 

\newpage

\section*{Appendix}
\subsection*{Calculation of Lyapunov spectra and Kolmogorov-Sinai entropy}

In this Appendix, we largely follow the corresponding section of Ref. \cite{Kanderschaos}. The largest Lyapunov exponent $\lambda_1$, describing the time-averaged rate of the strongest exponential separation of system trajectories in the tangent bundle, is used to determine whether a dynamical system is stable or chaotic. For $\lambda_1 < 0$, nearby trajectories converge, while $\lambda_1 > 0$ implies divergence of nearby trajectories and hallmarks chaos. At the critical point, $\lambda_1 = 0$; in its neighborhood, the system experiences a critical slowing down of the dynamics, where small perturbations can have long-lasting effects. We numerically determined $\lambda_1$ using the local linearization along the system's trajectory, i.e., the Jacobian matrix of the neural network \cite{Stoop1988, Peinke1992}. This powerful method not only yields $\lambda_1$, but provides the whole Lyapunov spectrum, i.e., all Lyapunov exponents of the system.

Every neuron lives in a three-dimensional state space: the two state variables $x^{(i)}_n$ and $y^{(i)}_n$, and the synaptic input variable $I^{(i)}_{n}$, into which also the temporally sparse external inputs are incorporated. The Jacobian matrix for a single neuron has the form

\begin{equation}
\centering
    J^{(i)}_n =
    \begin{dcases}
        \begin{bmatrix}
            \frac{\psi}{(1-x^{(i)}_n)^2} &1  &\beta \\ 
            -\mu &1  &\mu \\ 
            \Theta_n^{(i)} &0  &\eta 
        \end{bmatrix} & x_n^{(i)} \leq 0,\\
        \\
        \begin{bmatrix}
            0 &1  &\beta \\ 
            -\mu &1  &\mu \\ 
            \Theta_n^{(i)} &0  &\eta 
        \end{bmatrix} & 0 < x_n^{(i)} < \psi + u_n^{(i)} \wedge x_{n-1}^{(i)}\leq 0,\\
        \\
        \begin{bmatrix}
            0 &0  &0 \\ 
            -\mu &1  &\mu \\ 
            \Theta_n^{(i)} &0  &\eta 
        \end{bmatrix} & x_n^{(i)} \geq \psi + u_n^{(i)} \lor x_{n-1}^{(i)} > 0,\\
    \end{dcases}
\end{equation}
where $\Theta_n^{(i)} = -W(\sum^N_{j=1} w_{ij} \xi^{(j)}_n + w_{ext} \xi^{ext(i)}_n)$. The extension of the single neuron Jacobian matrix to the full network is straight-forward: the state variables of a neuron do not directly depend on the state variables of other neurons because the interaction is only through spike events and we can write the Jacobian of the full network, $J^{net}_n$, as a $3N \times 3N$ block diagonal matrix with the Jacobians of the individual neurons on the diagonal and all other elements being equal to 0.

Lyapunov exponents are obtained by following how a unit sphere $O_n$ is by the Rulkov Jacobian, into an ellipsoid $J^{net}_n O_n$.  A one-step growth rate of a unit length base vector $\mathbf{y}^{(d)}_n, \, d=1\ldots 3N$ is thus given by the length of the mapped vector $ \left \|\mathbf{y}^{(d)}_{n+1}\right \|$. By applying a Gram-Schmidt orthonormalization procedure,  a new unit sphere is reconstructed (with generally rotated base vectors) and the growth rates into their directions are determined anew. 
Maintaining the initial indexing of the unit base vectors and repeating this procedure, after $n$ iterations the separation of trajectories into the direction described by index $d$ is  
$r^{(d)}_n =  \left \| \mathbf{y}^{(d)}_1 \right \| \left \| \mathbf{y}^{(d)}_2 \right \| \hdots \left \| \mathbf{y}^{(d)}_n \right \|$. Owing to the Gram-Schmidt procedure, for $n$ large, index $d=1$ describes the largest and index $d=3N$ the smallest, separation in the tangent bundle. If we are interested in exponential separation $r^{(d)}_n = e^{\lambda_d n}$, the long-time $n\rightarrow \infty$  behavior of the system is described by the Lyapunov exponents 
\begin{equation}
\lambda_d = \lim_{n \rightarrow \infty} \frac{1}{n} \sum^n_{t=1}\text{ln}\left \| \mathbf{y}_t^{(d)} \right \|, \, d=1..3N,
\end{equation}
where the sign of the first exponent provides the information whether the system is 'chaotic' $(\lambda_1>0)$ or not $(\lambda_1\leq 0)$.
The orthonormalization procedure is computationally expensive, which makes the calculation of Lyapunov exponents for large networks slow.  We calculated the Lyapunov exponents of the subcritical, critical and supercritical networks for 10 random configurations out of the 50 that we used to obtain the avalanche size and lifetime distributions. The simulation length for the calculations was kept at $1.5 \cdot 10^5$ time steps. The Lyapunov exponents converged well, but, to take care of potential fluctuations, the final value of $\lambda_d$ was obtained by averaging over the last 5000 steps.

The spectrum of Lyapunov exponents yields directly the value of the the Kolmogorov-Sinai  entropy as the sum of the positive individual exponents. This value yields a more global picture than largest exponent alone and therefore is a helpful quantity for describing in particular how coherent chaos in the network at a given parameter is.

\subsection*{Results from network ensembles and varied node dynamics}
\begin{figure}[h]
\centering
\includegraphics[width=0.6\linewidth]{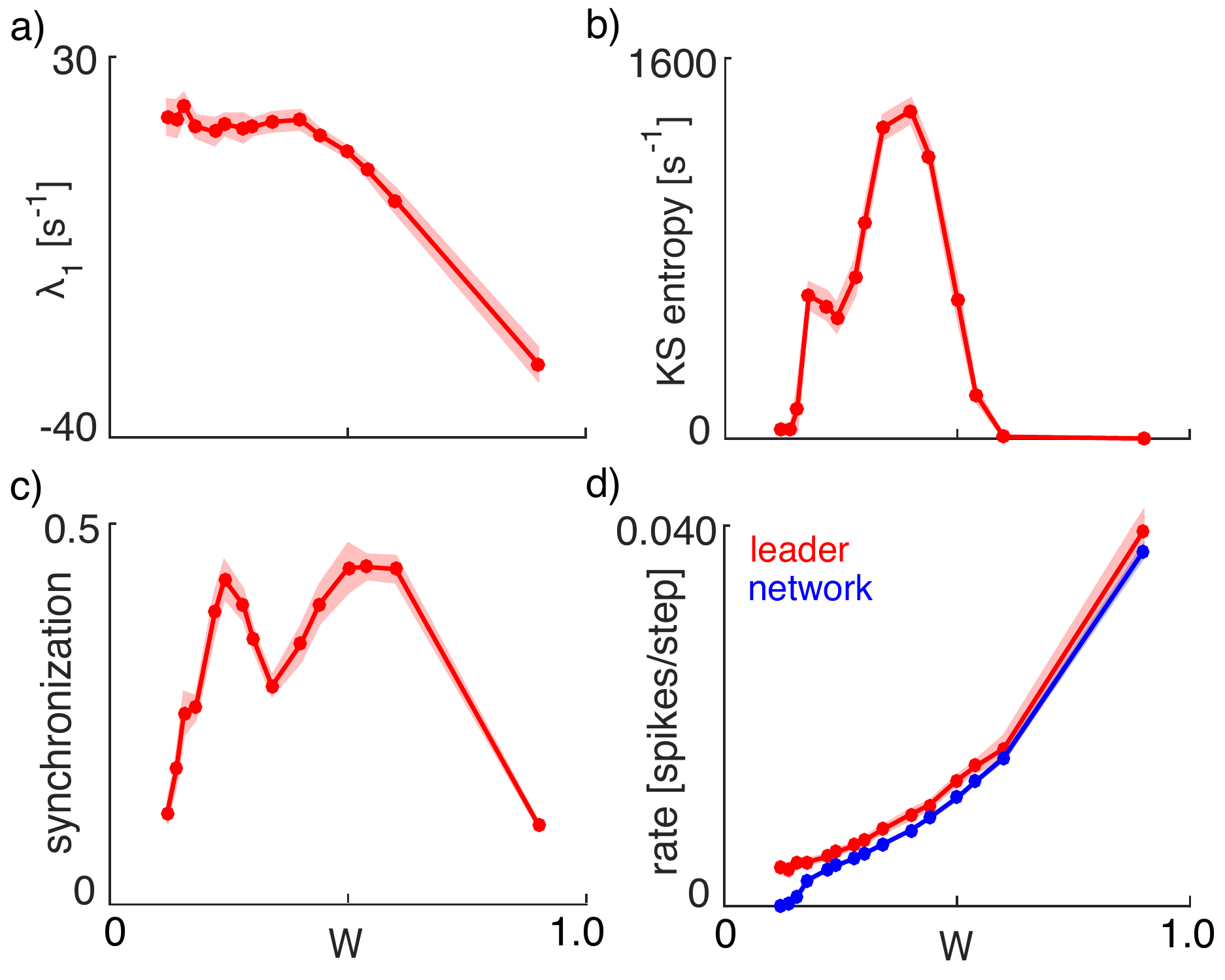}
\caption{Ensemble results: a) Largest Lyapunov exponent $\lambda_1$, b) Kolmogorov-Sinai entropy, c) synchronization measure, d) firing rates of the leader neuron (red) and the average network firing rate excluding the leader neuron (blue). Data points are results averaged over 10 network realizations; shaded area indicate one standard deviation around the mean. Each simulation for every value of W has a different network adjacency matrix and a different sampling of the random parameters.}
\label{ensemble_results}
\end{figure}

Ensembles based on randomized values of the network and neuron parameters, are largely similar to the characteristic network results (Fig. \ref{ensemble_results}). By parameter randomization, the synchronization measure emerged to be the most affected: Compared to the characteristic case, the first prominent peak of the synchronization measure (at $W \approx$ 0.24) is noticeably attenuated. In parallel, the first valley of the Kolmogorov-Sinai entropy and the Largest Lyapunov exponent become somewhat less pronounced. The general transition paradigm, as described in the main manuscript, remains, however, unchanged.

\begin{figure}[h]
\centering
\includegraphics[width=0.8\linewidth]{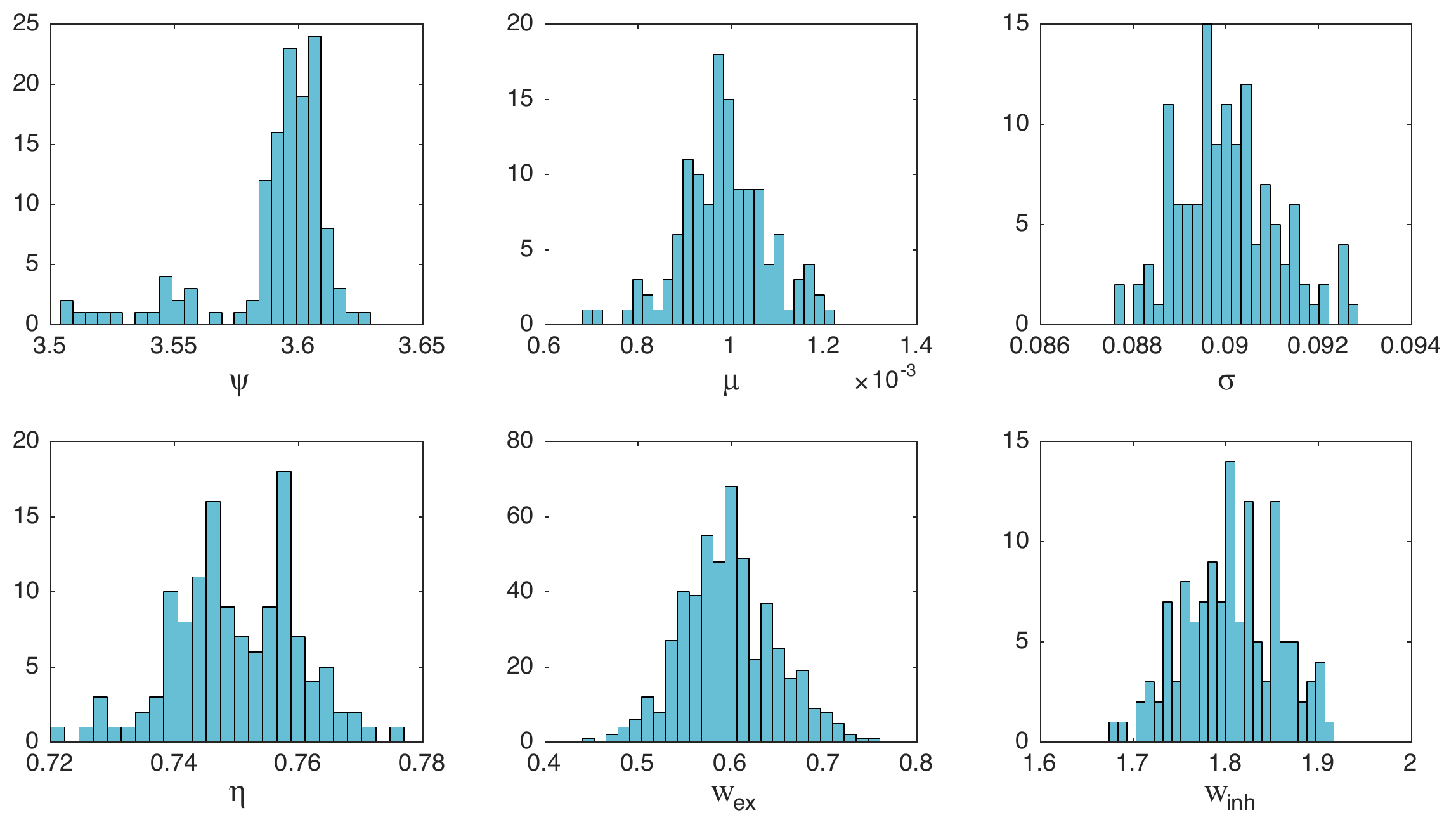}
\caption{Examples of the random samples drawn from the parameter distributions.}
\label{ensemble_parameters}
\end{figure}

The parameters were sampled from normal distributions $\mathcal{N}(mean,variance)$: $\sigma = \mathcal{N}(0.09,10^{-6})$, $\eta = \mathcal{N}(0.75,10^{-4})$, $w_{ex} = w_{ext} = \mathcal{N}(0.6,0.0025)$, $w_{inh} = \mathcal{N}(1.8,0.0025)$, $\mu = \mathcal{N}(0.001,10^{-8})$. ``Leader" neuron's parameter $\sigma$ was sampled from $\mathcal{N}(0.103,10^{-6})$. The parameter $\psi$ was sampled from a normal distribution $\psi = \mathcal{N}(3.6,10^{-4})$ but, to implement a larger variation among the intrinsically silent neurons, 20\% of the $\psi$ values were replaced by samples from a uniform distribution over the interval $\psi \in (3.5,3.6)$. Similar effect could have been achieved by sampling smaller $\sigma$ values. Randomization of the parameters $x_{rp}^{ex}$, $x_{rp}^{inh}$ and $\beta$ was not implemented, as they would increase the variation of the synaptic input, but this is already achieved by random synaptic weights and decay constant $\eta$. Example histograms of the sampled parameter values for one network realization are shown in Fig. \ref{ensemble_parameters}. While the variances of the neuron parameter distributions may appear small, the Rulkov's map is sensitive to small changes in the parameter values and, e.g., increasing the spread of $\psi$ or $\sigma$ distributions can give rise to additional intrinsically spiking neurons, which in our case would be undesirable.

To check the robustness of our observations, we confirmed that they can also be observed in network models that obey varied node dynamics. To this end, we replaced the more realistic Rulkov's dynamics by one-dimensional leaky integrate-and-fire neuron dynamics. In the latter case, the state variable $x_n$ features an exponential decay, a ``spike" threshold $\theta$, and an absolute refractory period of length $T_R$ time steps:

\begin{equation}
x^{(i)}_{n+1} = \begin{cases} 
a x^{(i)}_{n}  + I^{(i)}_{n} & x^{(i)}_n < \theta, \\ 
0 &x^{(i)}_k \geq \theta, k \in (n-T_R, n].
\end{cases}
\end{equation}
Synaptic input was implemented as

\begin{equation}
I^{(i)}_{n+1} = W \left ( \sum_{j=1}^N w_{ij} H(x_n^{(j)} - \theta) + \xi_n^{ext(i)} \right ),
\end{equation}
where $N$ is the network size, $H(\cdot)$ is the unit step function ($H(x) = 1$ if $x \geq 0; H(x) = 0$ if $x < 0$), and $\xi_n^{ext(i)}$ is an external Poisson input, similarly to the model described in the main part of the manuscript. Note that $I^{(i)}_n$ does not represent a separate state variable.
The network size and the network topology was as in the main manuscript. The excitatory weights were set to $w_{ij} = 1$, and inhibitory weights to $w_{ij} = -3$. One excitatory neuron was made to spike regularly (our ``leader neuron"), by adding $I_{leader} = 0.015$ to its synaptic input. The rest of the node parameters were identical for all neurons: $a = 0.995, T_R = 4, \theta = 2$. For each neuron, the external input probability was $p(\xi^{ext(i)}_n = 1) = 0.002$, at any time step.

\begin{figure}[!!!!!!!!!!!!!!!!!!!!h!!!!!!!!!!!!!!!!!!!!!]
\centering
\includegraphics[width=0.8\linewidth]{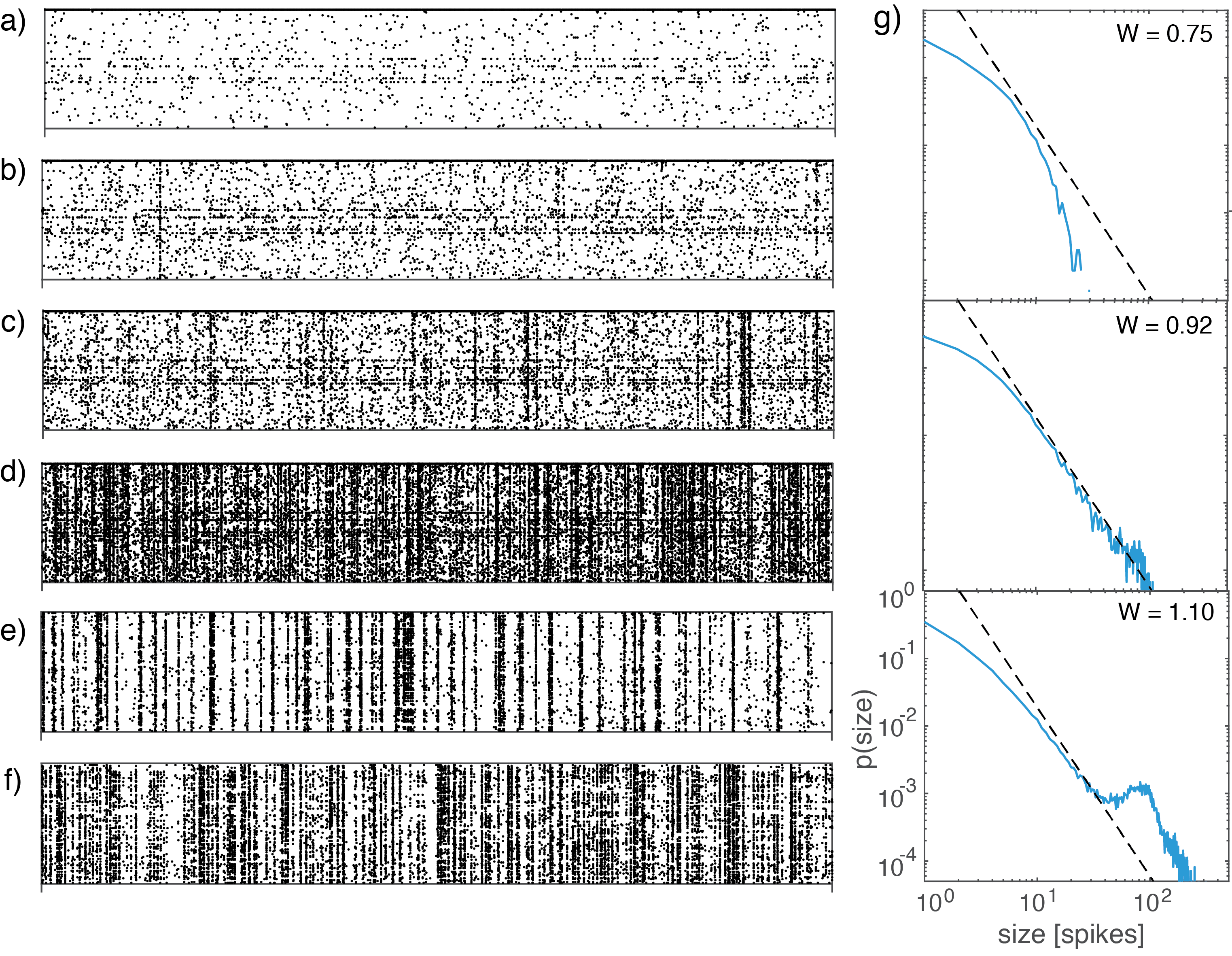}
\caption{Leaky integrate-and-fire network. a)-f) Raster plots for $W = 0.75, 0.85, 0.9, 1.0, 1.15, 1.7$ (top to bottom). Vertical axis: neuron number; horizontal axis: time (a)--d): 200000, e) --f): 20000 time steps). g) Avalanche size distributions. Dashed line: power-law with exponent $\alpha = 2.47$, from a fit at $W= 0.92$ (range of fit = $7 \ldots 70$; p-value = 0.054). Avalanches were obtained using temporal binning with a bin size equal to the average inter-event interval; distributions show results from pooled simulations of 10 network realizations}
\label{evolution_if}
\end{figure}

Similarly to our original model, the leaky integrate-and-fire network exhibits a critical point, where that avalanche size distribution has a power-law exponent of $\alpha \approx 2.5$ (Fig. \ref{evolution_if}, right-hand-side panels). The transition emerges at a changed value of $W$ because of the different sensitivity to inputs of the maps (the distance between resting state and spiking threshold is larger for our integrate-and-fire map than for Rulkov dynamics). Also, in the integrate-and-fire case, approaching the critical point is accompanied with the emergence of population-sized bursts (Fig. \ref{evolution_if}c)). These bursts are responsible for the persistent hump that develops at the tail of the distributions at larger values of $W$. As the weight scaling parameter is increased,
the network thus undergoes a structural transformation paradigm that has some similarity to that of the Rulkov dynamics case, albeit differences in the precise nature of these transformations could be expected.

\newpage

\section*{Declaration of conflict of interests }

The authors declare that the research was conducted in the absence of any commercial or financial relationships that could be construed as a potential conflict of interest.

\section*{Author Contributions}

KK and RS designed the research; KK, TL, YU and WHS performed calculations; RS, KK, and TL wrote the manuscript. All authors checked and approved the manuscript. 

\end{document}